\documentclass[]{ieeeconf}
\IEEEoverridecommandlockouts

\newcommand{\subparagraph}{}

\usepackage{graphicx,psfrag,dblfloatfix}
\usepackage{amsmath,amssymb,amsfonts,mathrsfs}
\usepackage{color,epstopdf}
\usepackage{algorithmic}
\usepackage{bm}
\usepackage{enumerate} 
\usepackage{subfigure}
\usepackage{float}
\usepackage{afterpage}
\usepackage[letterpaper,top=19.1mm,bottom=19.1mm,right=19.1mm,left=19.1mm]{geometry}
\usepackage{verbatim}
\usepackage{multicol,setspace}
\usepackage{titlesec}
\titlespacing\section{0pt}{12pt plus 0pt minus 2pt}{4pt plus 2pt minus 2pt}
\titlespacing\subsection{0pt}{8pt plus 0pt minus 2pt}{2pt plus 0pt minus 2pt}
\allowdisplaybreaks
\newtheorem{remark}{Remark}
\newtheorem{assumption}{Assumption}

\newtheorem{theorem}{Theorem}
\newtheorem{lemma}{Lemma}

\newtheorem{proposition}{Proposition}

\newcommand{\union}{\cup}

\newcommand{\Pcal}{\mathcal{P}}
\newcommand{\Qcal}{\mathcal{Q}}

\newcommand{\R}{\mathbb{R}}

\newcommand{\eps}{\varepsilon}
\newcommand{\V}{\mathcal{V}}    %for graphs

\newcommand{\rmd}{\text{d}}

\def\qedp{\hspace*{\fill}~{\tiny $\blacksquare$}}
\def\be{\begin{equation}}
\def\ee{\end{equation}}
\def\ba{\begin{array}}
\def\ea{\end{array}}
\def\bqa{\begin{eqnarray}}
\def\eqa{\end{eqnarray}}

\definecolor{darkgreen}{rgb}{0.0, 0.55, 0.0}
\definecolor{amaranth}{rgb}{0.9, 0.17, 0.31}
\usepackage[prependcaption,colorinlistoftodos]{todonotes}

\begin{document}
\setlength{\abovedisplayskip}{5pt}
\setlength{\belowdisplayskip}{5pt}

\title{Are energy savings the only reason for the emergence \\
	of bird echelon formation?}
\author{ Mingming Shi and Julien M. Hendrickx \vspace{-1.5em}
\thanks{M. Shi and J. M. Hendrickx  are with ICTEAM Institute, UCLouvain, Louvain-la-Neuve,  1348, Belgium. The work was supported by the ``RevealFlight'' Concerted Research Action (ARC) of the Federation Wallonie-Bruxelles.
	Email: {\tt\small mingming.shi@uclouvain.be, julien.hendrickx@uclouvain.be}
}
}

\maketitle

\begin{abstract}
We analyze the conditions under which the emergence of frequently observed echelon formation can be explained solely by the maximization of energy savings. We consider a two-dimensional multi-agent echelon formation, where each agent receives a benefit that depends on its position relative to the others, and adjusts its position to increase this benefit. We analyze the selfish case where each agent maximizes its own benefit, leading to a Nash-equilibrium problem, and the collaborative case in which agents maximize the global benefit of the group. We provide conditions on the benefit function under which the frequently observed echelon formations cannot be Nash equilbriums or group optimums.

We then show that these conditions are satisfied by the conventionally used fixed-wing wake benefit model. This implies that energy saving alone is not sufficient to explain the emergence of the migratory formations observed, based on the fixed-wing model. Hence, either non-aerodynamic aspects or a more accurate model of bird dynamics should be considered to construct such formations.
\end{abstract}

\section{introduction}

Formation control where multiple agents collaborate to move in certain shapes received extensive interest in the literature, see e.g., the survey \cite{oh2015survey}. 
Different methods based on available sensor measurements, e.g., position \cite{dong2008cooperative,ren2007distributed}, distance \cite{anderson2008rigid,hendrickx2007directed} and bearing \cite{zhao2015bearing,chen2020angle}, have been proposed to achieve formations for various agent dynamics. There is also a research line focusing on imitating the collective behavior of animals, e.g., birds flock or fish school, by designing simple local interaction rules \cite{reynolds1987flocks}-\nocite{vicsek1995novel,olfati2006flocking}\cite{tanner2007flocking}. Despite these fruitful results, the existing researches mostly focus on the actions agents take in order to form and maintain specific shapes, or on how phenomenological behavior models may result in formation-like behaviors. But, the benefits of these formations and their influence on the emergence of formations are rarely addressed.

In nature, several animal species, e.g., birds, fish and lobsters, move in specific formations that are argued to be caused by energy saving \cite{trenchard2016energy}. In particular, it is well-acknowledged that migrating birds adopt the eye-catching line formation because each follower bird reduces energy expenditure by exploiting the extra supportive lift from the wake of the front neighboring bird \cite{Heppner_Orginizeflight}-\nocite{weimerskirch2001energy}\cite{hummel1983aerodynamic}. By regarding birds as fixed wings, early researches \cite{hummel1983aerodynamic}-\nocite{badgerow1981energy,hummel1995formation}\cite{cutts1994energy} have tested the energy saving mechanism. Though the predicted relative position of neighboring birds is consistent with the observations of migrating birds, the position of each bird was always pre-fixed, without considering birds' incentive to pick the preferred position. Some papers in the last decade \cite{Cattivelli-Model,li2017v} also seek to construct line formations based on modified fixed wing models. However, their modification violates the wake evolution in aircraft experiments \cite{Greene-appmodel}, hence the conclusion could be questioned. Moreover, other non-aerodynamic factors are also considered in these work. Hence the actual emergence of the specific formation shapes (echelon or V) remains unexplained on multiple aspects, such as birds interests in energy saving, sensing ability, and action strategies. A first fundamental question is whether migrating formations emerge purely based on energy saving? To answer this, in \cite{Mingming_unpublished} we have recently tried employing the fixed-wing model to numerically constructing the echelon formation for birds by assuming all of them are either selfish or cooperative in energy optimization; see Section \ref{Sec:PreB} for more detailed explanation about these behaviors. Surprisingly, no observation-similar echelon formation has been found in any of the situations.

Our contribution in this paper is to theoretically confirm this result. We study the general two-dimensional multi-agent echelon formations based on benefit optimization. In our setting, each agent can receive from any other agent a benefit that depends on its relative position to that agent. A leader is fixed at the front of the group, while other followers can adjust their positions and their behaviors are purely guided by benefit optimization. Same as in our trial in constructing migratory formations, we consider that all agents are either selfish or cooperative, resulting in a self-benefit maximization non-cooperative game or a cooperative total benefit optimization problem, respectively. Related to the emergence of echelon formations, our focus is to derive conditions of the inter-agent benefit, under which there cannot exist a Nash equilibrium of the self-benefit game and/or the (local) optimum of the total benefit optimization, at which the relative position of each neighboring-agents lies within some proper set. 

This question is close to constrained non-cooperative games and maximization, where the existence of equilibriums or optimums could be guaranteed by requiring objective functions to be continuous 
 or concave \cite{bacsar1998dynamic,rosen1965existence}. But, unlike these problems, we focus on whether the unconstrained game or maximization has some equilibriums or optimums that are within the desired set by coincidence. 

We derive several results by analyzing the necessary condition of the existence of the Nash equilibrium and/or the maximum. Based on these results, we confirm the numerical results in \cite{Mingming_unpublished} using the fixed-wing model that birds behaving purely to maximize energy savings may not be sufficient to create the migratory formation.

%We note that the setting of the paper could also be applied to study other benefit induce formation in nature, for instance, the single-file queuing of spiny lobsters. 

The rest of the paper is organized as follows: In the next section, we first explain echelon formations, the benefit optimization problems with different birds interests and the considered Nash equilibrium and optimum. Then, we formulated the problem of interest. Section \ref{sec:NE} and \ref{Sec:CE} present conditions on the inter-agent benefit such that the considered Nash equilibrium and optimum, respectively, cannot exist. In Section \ref{Sec:Application}, we apply the proposed theoretical conditions to analyzing the fixed-wing wake model and justify our numerical results. At last, Section \ref{Sec:conclusion} concludes the paper and discuss the implication of the results. The Appendix provides the proof for an intermediate result in Section \ref{Sec:conclusion}. 

\section{Preliminaries}\label{Sec:profor}

\subsection{Notations}
Let $\text {Id}(\cdot)$ be the identity map.  For a negative interval $\Pcal \subset \R_-$, we denote by $-\Pcal$ and $2\Pcal$ the image of $-\text{Id}(a)$ and $2\text{Id}(a)$ when $a\in \Pcal$, respectively. 
 For a differentiable function $f(x): \R^m\rightarrow \R, m\ge 2$, we denote by $\frac{\partial f(x^*)}{\partial x_i}=\frac{\partial f(x)}{\partial x_i}\vert_{x=x^*}$ the partial derivative of $f(x)$ with respect to the $i$th component of the argument at $x^*\in \R^m$. Moreover, if $m=2$, we denote $f_x(a,b) = \frac{\rmd f(x,b)}{\rmd x}\vert_{x=a}$, with $a, b\in \R$ before Section \ref{Sec:Application}.

\subsection{Echelon formation, agents benefits and interests}\label{Sec:PreB}
We consider $n+1$ agents with one leader with label 0, and $n\ge 2$ followers labeled from 1 to $n$. Let $V=\{1,...,n\}$ denote the set of followers. For each agent $i\in V\union\{0\}$, we call agent $j$, with $j=i \mp k\ge 0$ with $1\le k\le n$, the $k$-hop front (back) neighbor of $i$. %The front or back neighbor is a short for the 1-hop corresponding neighbor. 
${\mathcal N}_{i}\subset V\union \{0\}$ denotes the set of $1$- and $2$-hop neighbors of $i$. 
Each agent $i\in V\union\{0\}$ has a position $p_i=[x_i \ y_i]^\top\in \R^{2}$ and specifically $p_0=0_{2}$ in this paper. Let $X = [x_1\ ... \ x_n]^\top$, $Y=[y_1\ ... \ y_n]^\top$ and $p=[X^\top\ Y^\top]^\top$. In the later, the $x$ and $y$ directions are also called the longitudinal and lateral directions, respectively. Backward motion means moving at the negative $x$ direction. Let $p_{ij}=p_i-p_j$ and $x_{ij}=x_i-x_j$ for any unequal $i, j\in V\union \{0\}$. 

Each agent $i\in V\union\{0\}$ can gain a benefit $f^{(i)}(p)$ that depends on agents positions $p$, from all others. Consistently with the analysis of bird formations \cite{Heppner_Orginizeflight}, where the energy saving of the bird is additive and each bird is affected mostly by two front and back birds, we assume that $f^{(i)}(p)$ can be decomposed as the sum of the benefits $f(p_{ij})$ agent $i$ gets from $j\in \mathcal N_i$, where $f(\cdot): \R^2\rightarrow \R$ is the inter-agent benefit. Mathematically,
\begin{eqnarray}\label{eq:fi(p)}
f^{(i)}(p)=\sum_{j\in {\mathcal N}_i} f(p_{ij})=\sum_{j\in V\union \{0\}}^{|i-j|\le 2,\ i\ne j} f(p_{ij})
\end{eqnarray}

The total benefit $J(p)$ of the group is then the sum of the benefit of all agents:
\begin{align} \label{eq:Jp}
J(p)&=\sum_{i\in V\union\{0\}} f^{(i)}(p) =\sum_{i\in V\union\{0\}}\sum_{j\in {\mathcal N}_i } f(p_{ij})
%&=\frac{1}{2}\sum_{i,j\in V\union \{0\}}^{|i-j|\le 2,\ i\ne j} (f(p_{ij})+f(p_{ji}))
\end{align}

\begin{figure}[!t]
	\centering
	\subfigure[]{\includegraphics[scale =0.35]{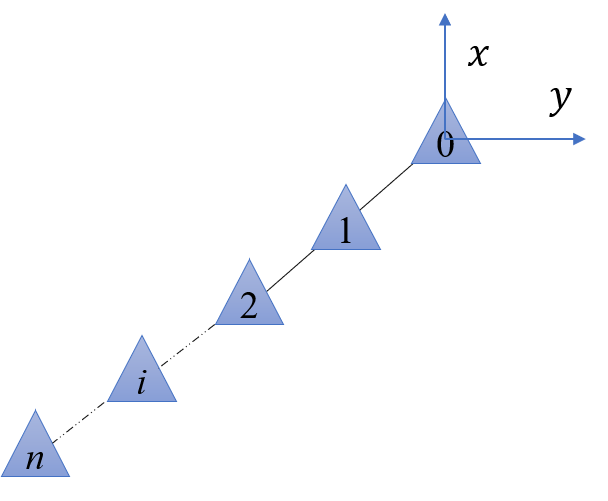}
		\label{fig.agentsformation}}
	\hfill
	\subfigure[]{\includegraphics[scale =0.35]{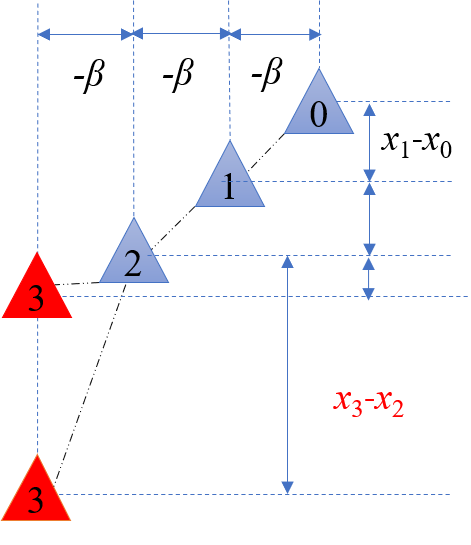}
		\label{fig.weirdformation}}
	\caption{Echelon formation and formations that are weird. (a) $n+1$ agents create an echelon formation in a plane. (b) Weird formation that should be excluded. Agent $3$ in red color is either too close or too far from the front neighbor longitudinally. }
	\label{fig.echelonformation}
\end{figure}

We focus on echelon formations as shown in Fig. \ref{fig.agentsformation}, where agents are aligned diagonally behind the leader in one side with equal neighboring-agents distance.  Motivated by the line formation of migrating birds where neighboring birds' lateral distances are almost the same but longitudinal distances are varied within proper range \cite{Heppner_Orginizeflight,hummel1983aerodynamic},  we allow the formation to be deviated from the strict echelon shape. Specifically, we focus on the left echelon formation where the position $p_i$ of each follower $i\in V$ satisfies 
\begin{align}
y_i = -i\beta, \quad x_{i(i-1)}\in \Pcal
\end{align}
where $\beta$ is a positive and $\Pcal = [-\alpha_l,-\alpha_s]$ with $\alpha_s,\alpha_l$ two preset positives satisfying $\alpha_s\le \alpha_l$. Let $Y(\beta) = [-\beta\ ...\ -n\beta]^\top$. Then $f^{(i)}(p)$ and $J(p)$ can also be denoted as $f^{(i)}(X,Y(\beta))$ and $J(X,Y(\beta))$, respectively.

In the paper, we fix $\beta$ and consider to construct the echelon formation of interest by assuming that followers can adjust their longitudinal position $x_i$ based on benefit maximization. Two different agent attitudes are considered. In the first, all followers are selfish and would like to maximize their own benefits $f^{(i)}(p)$. This leads to a non-cooperative game and we are interested in the Nash equilibrium (NE) of the longitudinal positions, which is defined as the vector $X^*=[x^*_1\ \cdots \ x^*_n]^\top\in \R^n$  satisfying the condition below for each $i\in V$:
\begin{align}
f^{(i)}(x_i^*,x_{-i}^*,Y(\beta)) \le f^{(i)}(x_i,x_{-i}^*,Y(\beta)), \quad \forall x_i\in \R
\end{align}
where $x^*_{-i} = [x_1^*\ \cdots\ x_{i-1}^{*}\ x_{i+1}^*\ \cdots\ x_n^*]^\top$. The NE, if exists, corresponds to agents longitudinal positions with the property that no agent can increases its own benefit by choosing a different position unilaterally. 

In the second, all agents cooperative to maximize the group total benefit $J$ %in \eqref{eq:Jp}
 and we are interested in the cooperative equilibrium (CE), which is the vector of agents' longitudinal positions $\bar X^*=[\bar x^*_1\ \cdots \ \bar x^*_n]^\top\in \R^n$ that reaches a local maximum of $J$.
\begin{align}\label{eq:cooperequi}
\bar X^* := \arg\max_{X\in  \mathcal B_{X}} J(p)=\arg\max_{X\in  \mathcal B_{X}} J(X,Y(\beta))
\end{align}
where $\mathcal B_{X}$ is a neighborhood of $X$. It is proper to consider the local maximum since without prior knowledge of the global maximum, the cooperative followers have no incentive to shift away from a local maximum. Moreover, the global maximum is also a local maximum, thereby satisfies \eqref{eq:cooperequi}.

 We denote $p_i^* = [x_i^* \ -i\beta]^\top$, $\bar p_i^* = [\bar x_i^* \ -i\beta]^\top$, $p^*=[(X^*)^\top\ Y(\beta)^\top]^\top$, and $\bar p^*=[(\bar X^*)^\top\ X(\beta)^\top]^\top$, respectively. Consistently with the considered echelon formation, we only focus on the NE $X^*$ and CE $\bar X^*$ with $x^*_{i(i-1)}\in \Pcal, i\in V$ and $\bar x^*_{i(i-1)}\in \Pcal, i\in V$, respectively, for a negative closed interval $\Pcal$ and positive $\beta$, which can be preset based on practical tasks or observations. %, e.g., bird formations. 
Whether a considered echelon formation can be constructed based on benefit maximization should relate to if there exist the equilibrium of interest.

\subsection{Problem formulation}

In view of $f^{(i)}(p)$ and $J(p)$, % in \eqref{eq:fi(p)} and \eqref{eq:Jp},
the existence of  $X^*$ and $\bar X^*$ should depend on the properties of the inter-agent benefit $f$.
By imposing strong concavity \cite{rosen1965existence} on the benefit within the interval $\Pcal$, it might not be difficult to obtain conditions that guarantee the existence of the equilibrium of interest. While, in our efforts to reconstruct the migratory formation of birds based purely on benefit maximization, no equilibrium that corresponds to observation-similar echelon formation has been found. To explain this, we focus on the following problem in the paper.

\emph{Problem 1.} Given $n\ge 2$ agents, an interval $\Pcal = [-\alpha_l,-\alpha_s]$ with $0<\alpha_s\le \alpha_l$ and a $\beta>0$, under what conditions on $f$, the NE $X^*\in \R^n$ with $x^*_{i(i-1)}\in \Pcal, i\in \V$ and/or the CE $\bar X^*\in \R^n$ with $\bar x^*_{i(i-1)}\in \Pcal,i\in V$ are impossible.

 %The benefit function could be any type, and even discontinuous or discrete. 
 At this stage, we impose the following assumption on the inter-agent benefit  $f$, allowing to consider its derivative.
\begin{assumption} 
(a) $f(x,y)$ with $x,y\in \R$ is continuous in $\R^2$ and is continuously differentiable when $x\ne 0$ \label{Asmp: f(p)diff} (b) $f(x,y)=f(x,-y)$ for $x,y\in \R$. \label{Asmp: f(p)diff2}
\end{assumption} 

The continuity of the derivative of $f$ is not assumed at $x=0$ since the inter-agent benefit may have an acute change when an agent shifts from the back to the front of another agent longitudinally.

	%no issue should arise if we consider $\frac{\rmd f(x,y)}{\rmd x} $ for $x\in \Pcal\cup 2\Pcal \cup -\Pcal \cup -2\Pcal$. 

Note that $f(p_{ij})$ takes $p_{ij}=[x_{ij}\ y_{ij}]^\top$ and remember the definition of the NE in \eqref{eq:cooperequi}. Then by the chain rule, if a NE $X^*$ with $x^*_{i(i-1)}\in \Pcal, i\in V$ exists, it should satisfy the equation below for each $i\in V$
\begin{align} \label{eq:selfish_opt}
0 = \frac{\partial f^{(i)}(p^*)}{\partial x_i} =\sum_{j\in {\mathcal N}_i} \frac{\partial f(p_{ij})}{\partial x_{ij}}\frac{\partial x_{ij}}{\partial x_i}\biggr\vert_{p^*} 
= \sum_{j\in {\mathcal N}_i}\frac{\partial f(p_{ij}^*)}{\partial x_{ij}}
\end{align}
By contrast, if a CE $\bar X^*$ with $\bar x^*_{i(i-1)}\in \Pcal, i\in V$ exists, it should satisfy the following equation for each $i\in V$
\begin{align} \label{eq:group_opt}
0= \frac{\partial J(\bar p^*)}{\partial x_i}%&= \frac{ \sum_{k\in V\union\{0\}}\sum_{j\in {\mathcal N}_k} \partial f(p_{kj})}{\partial x_i}\biggr\vert_{p=\bar p^*}\nonumber\\
&= \sum_{k\in V\union\{0\}}\sum_{j\in {\mathcal N}_k} \frac{\partial f(p_{kj})}{\partial x_{kj}}\frac{\partial x_{kj}}{\partial{x_i}}\biggr\vert_{\bar p^*}\nonumber\\
 &= \sum_{j\in {{\mathcal N}_i}} \frac{\partial f(\bar p_{ij}^*)}{\partial x_{ij}}-\frac{   \partial f(\bar p_{ji}^*)}{\partial x_{ji}}
\end{align}
where the last equality is from \eqref{eq:fi(p)} and \eqref{eq:Jp}.

Hence, checking if the equilibriums of interest exist is equivalent to testing if there exist a solution $X^*$ with $ x_{i(i-1)}^*\in \Pcal, i\in V$ to equation \eqref{eq:selfish_opt} and/or a solution  $\bar X^*$ with $\bar x_{i(i-1)}^*\in \Pcal,i\in V$  to \eqref{eq:group_opt}, respectively.

\section{Nonexistence of the NE of interest}\label{sec:NE}
In this section, we focus on the selfish case and discuss the conditions on $f$ such that there exists no NE of interest.

\subsection{Three agents case}

We first consider the simple case of three agents, 1 leader and 2 followers and $V=\{1,2\}$). 
  Intuitively, if the increment of agent 1's benefit from its back neighbor (agent 2) is more than its benefit loss from the front neighbor (agent 0) when agent 1 moves backward, then agent 1 would like to move backward to get more benefit. If this always holds when $x_{21},x_{10}\in \Pcal$, then agent $1$ cannot be static when $x_{21},x_{10}\in \Pcal$. In other words, the NE of interest cannot exist. 
The theorem below formulates this intuition.
\begin{theorem}\label{thm:noeo_3agents}
	For $n=2$, $\beta>0$, a closed interval $\Pcal\subset \R_-$, if $f(\cdot)$ satisfies Assumption 1 and 
	\begin{align}
\max_{x\in -\Pcal }f_x(x,-\beta)<-\max_{x\in \Pcal }f_x(x,-\beta)\label{eq:thm1}
\end{align}
 then there exists no NE $X^*\in \R^2$ with $x^*_{10},x^*_{21}\in \Pcal$.
\end{theorem}

\emph{Proof.} Suppose the conclusion is incorrect and there exists an NE $X^*\in \R^2$ satisfying $x_{10}^*,x_{21}^*\in \Pcal$. Since $f$ is continuously differentiable when $x\ne 0$ by Assumption 1(a), $X^*$ should satisfy \eqref{eq:selfish_opt}. 

Consider $i=1$ and notice that ${\mathcal N}_1=\{0,2\}$ and $p_i^*=[x_i^*\ -\beta]^\top$, 
%and ${\mathcal N}_2=\{1,2\}$ and $p_i^*=[x_i^*\ \text{--}i\beta]^\top$, these two equations above lead to
\eqref{eq:selfish_opt} leads us to
\begin{align*}
f_x(x_{10}^*,-\beta)+f_x(x_{12}^*,\beta) =0 
\end{align*}
%where the first term in the left side of the equality is due to  $x_{10}^*=x_1^*-x_0=x_1^*$. 
By $x^*_{10}\in \Pcal$, $x_{12}^*\in -\Pcal$ and Assumption 1(b), condition \eqref{eq:thm1} leads us to $ f_x(x_{12}^*,\beta)<-f_x(x_{10}^*,-\beta)$, contradicting the equality above. Hence, there exists no such $X^*$. \qedp

Theorem 1 is based on the analysis of agent 1. If we consider benefit change of both agents 1 and 2, a different result can be obtained. Consider just agent 1 and 2, namely, $f^{(2)} (p)=f(p_{21})$. If $f(x,-\beta)$ peaks  at $x=-\alpha$, then agent 2 should be $\alpha$ behind agent 1. Now take the leader 0 into account, namely $f^{(2)} (p)=f(p_{21})+f(p_{20})$. If %the benefit of an agent from its 2-hop neighbors 
$f(p_{20})$ changes very little when agent $2$ moves along the longitudinal direction, then the best $x_{21}$ that maximizes $f^{(2)}(p)$ would deviate very little from $-\alpha$. Hence when agent 1 moves longitudinally, if agent 2 wants to maximize the benefit, it should also move such that $x_{21}$ is within a very narrow interval $\Qcal\ni -\alpha$. Suppose this is true and $x_{12}\in -\Qcal$ always holds. Now assume that the increment of agent 1's benefit from agent 2 is more than the decrement of its benefit from agent 0 when agent 1 moves backward but keeps $x_{10}\in \Pcal$, %($x_{12}\in [\alpha'_s,\alpha'_l]$ always holds),
then agent $1$ would like to move backward, until $x_{10}\notin -\Pcal$.
This implies that there exists no NE $X^*$ with $x_{i(i-1)}\in \Pcal, i\in \V$.
This analysis can be formulated as another result, whose rigorous presentation relies on an assumption and several notations in the following.

%First, for the negative interval $\Pcal=[-\alpha_l,-\alpha_s]$ and positive $\beta$, we require $f(x.-\beta)$ to have a (local) maximum within the interval $\Pcal$. Note if this is not true, the simple case of two-agent formation where the follower's relative longitudinal position to the leader within the interval $\Pcal$ cannot exist. 
%\begin{assumption}\label{asmp:monotonicitydfpbeta}
%	There exists a critical point $-\alpha\in \Pcal$ such that $f(x,-\beta)$ is strictly increasing when $x\in [-\alpha_l, -\alpha)$ and strictly decreasing when $x\in (-\alpha, \alpha_s]$. Moreover, $f(x,-\beta)$ is strictly decreasing when $x\in -\Pcal$.
%\end{assumption}

%First, we make the following assumption.
\begin{assumption}\label{asmp:monotonicitydfpbeta}
(a) $f(x,-\beta)$ has a global maximum $-\alpha$, and is strictly increasing when $x<-\alpha$ and strictly decreasing when $x>-\alpha$. (b)The global maximum is in the interval of interest, $-\alpha \in \Pcal$.
\end{assumption}

 Assumption 2(a) can be regarded as the attribute of the benefit $f$. It is mild and satisfied by at least the benefit considered in Section \ref{Sec:Application}. Assumption 2(b) relates to the choice of the interval $\Pcal$. It is reasonable since otherwise a trivial conclusion could be obtained for the case of two agents that the follower 1 would never stay statically behind the leader 0, with $x_{10}\in \Pcal$.
 
 \begin{figure}[!t]
 	%		\subfigure[$\delta_1$ and $\delta_2$.]{\includegraphics[scale=0.38]{fun_exp_ee1.png}
 	%			\label{fig:fun_exmp_ee1}}
 	%		\hfill
 	\includegraphics[scale=0.42]{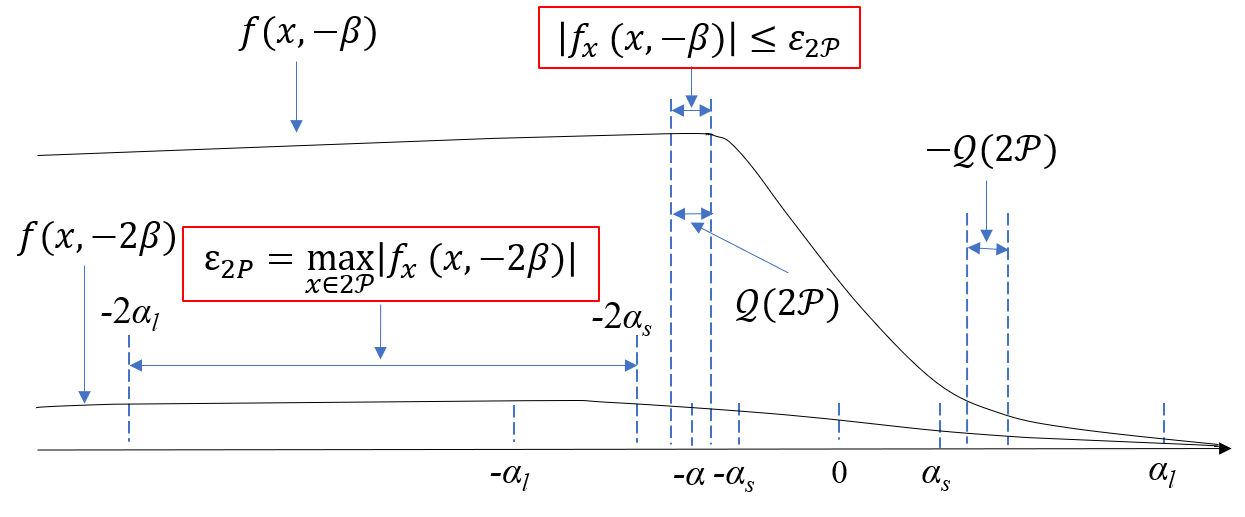}
 	\caption{Illustration of $f$, $\eps_{2\Pcal}$ and $\Qcal(2\Pcal)$, where $f(x,-2\beta)$ is flat in $2\Pcal$.}
 	\label{fig:fun_exmp_ee2}
 \end{figure}

We then characterize the narrow interval around $-\alpha$ mentioned  in the intuitive analysis before Assumption \ref{asmp:monotonicitydfpbeta}. For any non-empty closed interval ${\mathcal I}$, we denote
\begin{align}
\eps_{\mathcal I}:=\max_{x\in {\mathcal I}}\left| f_x(x,-2\beta)\right|\label{eq:eps1}
\end{align}	
and let
\begin{align}\label{eq:<=eps}
\Qcal(\mathcal I)= \{x\in \Pcal|\left|f_x(x,-\beta)\right|\le \eps_{\mathcal I}\}
\end{align}	
When $\mathcal I=2\Pcal$, $\Qcal(\mathcal I)$ relates to the narrow interval around $-\alpha$, though it may cover that.  See Fig. \ref{fig:fun_exmp_ee2} to get some vision of $\Qcal(2\Pcal)$. Formalizing the intuitive analysis, we have the following result.
\begin{theorem}\label{thm:noeo2_3agents}
For $n=2$,  $\beta>0$ and $\Pcal=[-\alpha_l,-\alpha_s]$ with $0<\alpha_s\le \alpha_l$, assume that $f$ satisfies Assumption 1  and \ref{asmp:monotonicitydfpbeta}. If 
	\begin{align}
	\max_{x\in -\Qcal(\mathcal I)} f_x(x,-\beta) <-\max_{x\in \Pcal }f_x(x,-\beta)\label{eq:thm2}
	\end{align}
with $\mathcal I = 2\Pcal$, then there exists no NE $X^*\in \R^2$ with $x^*_{10},x^*_{21}\in \Pcal$.
\end{theorem}

\emph{Proof.} First, the set $\Qcal(2\Pcal)$ is non-empty. In fact, since the maximum $-\alpha\in \Pcal$ and $f_x$ is continuously for $x\ne 0$ by Assumption 1 and \ref{asmp:monotonicitydfpbeta}, for any small $\eps_{\mathcal I}>0$, there should exist a neighborhood $\mathcal B_{-\alpha}$ of $-\alpha$ such that $|f_x(x,-\beta)|\le \eps_{\mathcal I}$ for all $x\in  \mathcal B_{-\alpha}$. We always have $\Qcal(\mathcal I)\supseteq\mathcal B_{-\alpha}\cap \Pcal$. %\blue{Note that $\Qcal$ could contain multiple isolated closed intervals and if it only contains one closed interval, it should be a neighborhood of $-\alpha$.}

Then, suppose there exists the NE $X^*\in \R^2$  of interest, a point satisfying \eqref{eq:selfish_opt}.  Notice that $\mathcal N_1=\{0,2\}, \mathcal N_2 = \{0,1\} $ and $p_i^*=[x_i^*\ -i\beta]^\top$,  hence \eqref{eq:selfish_opt} becomes
	\begin{align}
	0 & =  f_x(x_{10}^*,-\beta)+f_x(x_{12}^*,\beta) \label{eq:dfx_1}\\
	0 & = f_x(x_{20}^*,-2\beta)+f_x(x_{21}^*,-\beta)  \label{eq:dfx_2}
	\end{align} 
Since $x_{10}^*,x_{21}^*\in \Pcal$, $x_{20}^*=x^*_{10}+x^*_{21}\in  2\Pcal$, then from \eqref{eq:eps1}, $|f_x(x_{20}^*,-2\beta)|\le \eps_{2\Pcal}$. From equation \eqref{eq:dfx_2}, we also have $|f_x(x_{21}^*,-\beta)|\le \eps_{2\Pcal}$. Hence $x_{21}^*\in \Qcal(2\Pcal)$, or $x_{12}^*\in -\Qcal(2\Pcal)$. Then by \eqref{eq:thm2} and Assumption 1(b), $f_x(x_{12}^*, \beta)< -f_x(x_{10}^*,-\beta)$. However, this leads to a contradiction since the left side of \eqref{eq:dfx_1} would be negative. \qedp

In Theorem \ref{thm:noeo2_3agents}, condition \eqref{eq:thm2} should be satisfied for $\mathcal I=2\Pcal$.  Based on the analysis before Assumption \ref{asmp:monotonicitydfpbeta}, it may implicitly require the variation of $f(x,-2\beta)$ for $x$ in entire $2\Pcal$ to be small. For those inter-agent benefits $f$ that do not satisfy this condition, we can have another result if $f$ additionally satisfies Assumption 3 as follows.
\begin{assumption}\label{asm:dfp2beta}
	The benefit function $f(x, -2\beta)$ is strictly decreasing for $x \ge -2\alpha$.
\end{assumption}

% which implies that the benefit of agent 2 from the leader grows before $x_{20}$ decreases to exceed $-2\alpha$. This enforces agent 2 to be more than $2\alpha$ behind the leader longitudinally. In fact, if this is not true and $-2\alpha <x_{20}<-2\alpha_s$, then agent 2 should be behind agent 1 with $x_{21}<-\alpha$ to maximize its benefit, since the benefit of agent 2 from agent 1 also increases before $x_{21}$ decreases to $-\alpha$, according to Assumption 2. Since $x_{10}=x_{20}-x_{21}$, $x_{10}>-\alpha$ should hold. However, to maximize its benefit, agent 1 cannot be static if its benefits received from the leader and agent 2 both increase when agent 1 moves backward. Hence $x_{20}<-2\alpha$ holds. Now suppose this is true and
%agent 1's benefit from the leader changes very little when $x_{20}$ varies, then same as the analysis before Theorem \ref{thm:noeo2_3agents}, we would also have that no matter how agent 1 moves longitudinally, $x_{21}$ should belong to a very narrow interval $\Qcal\ni -\alpha$ to maximize $f_2(p)$. If the same conditions on the benefit function, mentioned in the paragraph before Theorem \ref{thm:noeo2_3agents}, holds, then there should not exist any NE of interest. 
%
%Based on these, we give a final result by exploiting all the assumptions at the beginning of this section.

%We then give the following result.
\begin{theorem}\label{thm:noeo3_3agents}
	For $n=2$, $\beta>0$ and $\Pcal=[-\alpha_l,-\alpha_s]$ with $0<\alpha_s\le \alpha_l$, assume that $f(\cdot)$ satisfies Assumption 1, \ref{asmp:monotonicitydfpbeta} and \ref{asm:dfp2beta}, If \eqref{eq:thm2} holds with $\mathcal I = [-2\alpha_l,-2\alpha]$, then there exists no NE $X^*\in \R^2$ with $x^*_{10},x^*_{21}\in \Pcal$.
\end{theorem}

\emph{Proof.} Suppose there exists an NE $X^*\in \R^2$ with $x_{10}^*,x_{21}^*\in \Pcal$. Then, $x_{20}^*\in  2\Pcal$, and $X^*$ should satisfy equation \eqref{eq:dfx_1} and \eqref{eq:dfx_2}. We consider two cases as follows.

\emph{Case 1. $x_{20}^*\in  [-2\alpha_l,-2\alpha]$.} Note $\eps_{\mathcal I}$ in \eqref{eq:eps1} is defined on $\mathcal I= [-2\alpha_l,-2\alpha]$ in this theorem. Following a similar argument as in the proof of Theorem \ref{thm:noeo2_3agents}, we have that the NE $X^*$ of interest cannot exist.

\emph{Case 2. $x_{20}^*\in  (-2\alpha,-2\alpha_s]$.} By Assumption \ref{asm:dfp2beta} and $2\alpha<\bar \alpha$, $f_x(x_{20}^*,-2\beta)< 0$. Then from \eqref{eq:dfx_2}, $f_x(x_{21}^*,-\beta)> 0$, implying $x_{21}^*<-\alpha$ according to Assumption \ref{asmp:monotonicitydfpbeta}. This, along with $x_{20}^*\in (-2\alpha,-2\alpha_s]$, leads to $x_{10}^*>-\alpha $ or $x_{10}^*\in (-\alpha,-\alpha_s]$. 
Hence $f_x(x_{10}^*,-\beta)<0$ according to Assumption \ref{asmp:monotonicitydfpbeta}. However, we also have $f_x(x_{12}^*,\beta)<0$ by Assumption 1(b) and \ref{asmp:monotonicitydfpbeta}. Hence, the left side of equation \eqref{eq:dfx_1} is less than zero. This leads to a contradiction. \qedp

\begin{remark}
	 Assumptions 1, 2 and 3 could be weakened. First, since the interval $\Pcal$ is finite, the conditions on $f$ in these assumptions could be imposed just for sets that cover all the intervals concerned. 
	 For instance, in Assumption 1(a) one could only require $f$ to be continuously differentiable in the set $(2\Pcal\cup\Pcal\cup-\Pcal\cup-2\Pcal)\times \R$. And in Assumption 2 (a), requiring $f_x(x,-\beta)$ to monotonically increase in $[-\alpha_l,-\alpha)$ and decrease in $(-\alpha,-\alpha_s]\cup -\Pcal$ is sufficient to draw the conclusion of Theorem \ref{thm:noeo2_3agents} and Theorem \ref{thm:noeo3_3agents}.
	  Second, the conclusion of Theorem \ref{thm:noeo2_3agents} would  still hold if Assumption 2(b) is discarded. In that situation, $\Qcal(2\Pcal)$ may be empty.  But this is not a problem since it implies a trivial case that \eqref{eq:dfx_2} is not satisfied.
\end{remark}
\smallskip

There is no strict advantage of using one theorem over others. On the one hand, by \eqref{eq:eps1} and \eqref{eq:<=eps}, $\Qcal(\mathcal I)\subseteq \Qcal(2\Pcal)\subseteq \Pcal$ with $\mathcal I=[-2\alpha_l,-2\alpha]$. Hence, condition \eqref{eq:thm2} in Theorem \ref{thm:noeo3_3agents} is easier to satisfy than that in Theorem \ref{thm:noeo2_3agents}, and than  condition \eqref{eq:thm1} in Theorem \ref{thm:noeo_3agents}. In other words, there may exist the benefits $f$ such that \eqref{eq:thm2}  is satisfied, but not \eqref{eq:thm1}. On the other hand, Theorems \ref{thm:noeo2_3agents} and \ref{thm:noeo3_3agents} require more knowledge and assumptions on $f_x$ than Theorem \ref{thm:noeo_3agents}, which may not be satisfied by the benefit $f$. 	In addition, unless $\eps_{\mathcal I}$ with $\mathcal I=2\Pcal$ or $\mathcal I =[-2\alpha_l,-2\alpha]$ is much small, 	$\Qcal(\mathcal I)$ may not be a small sub-set of  $\Pcal$ such that $\max_{x\in -\Qcal(\mathcal I)}f_x(x,-\beta)$ is much less than $\max_{x\in -\Pcal}f_x(x,-\beta)$. In that case, Theorems \ref{thm:noeo2_3agents} and Theorem \ref{thm:noeo3_3agents} may be not more useful than Theorem \ref{thm:noeo_3agents}.
	
%\begin{remark}
%	The key of Theorem \ref{thm:noeo2_3agents} and \ref{thm:noeo3_3agents} is to enforce $x_{21}$ within a small sub-set $\Qcal$ of $ \Pcal$.  The former achieves this by requiring implicitly a small upper bound on $|f_x(x,-2\beta)|$ for $x\in 2\Pcal$, which may be difficult to satisfy, while the later assumes that the upper bound on $|f_x(x,-2\beta)|$ is small for a sub-interval of $2\Pcal$, but requires the monotonicity of $ f(x,-2\beta)$.
%\end{remark}

\subsection{General case}
The following results extend Theorems \ref{thm:noeo_3agents}, \ref{thm:noeo2_3agents} and \ref{thm:noeo3_3agents} to the multiple agents case. Their proofs can be obtained by similar arguments in the previous subsection, and thereby are omitted here for space reason.
\begin{proposition}\label{prop:gcee1}
	For $n\ge 3$, a closed interval $\Pcal\subset \R_-$ and a $\beta>0$, assume $f(\cdot)$ satisfies Assumption 1. If $\max_{x\in -\Pcal }f_x(x,-\beta)<-\max_{x\in \Pcal }f_x(x,-\beta)-\eps_{2\Pcal}$, then there exists no NE $X^*\in \R^n$ with $x_{i(i-1)}^*\in \Pcal$ for each $i\in V$.
\end{proposition} 

%\emph{Proof.}  Suppose there exists a RNE $x^*\in \R^n$, then  $x^*$ should satisfy equation \eqref{eq:selfish_opt} for each $i\in V$.
%By $p_i^*=[x_i^*\ \text{--}i\beta]^\top$ and the fact that ${\mathcal N}_i$ contains the 1- and 2-hop neighbors of agent $i$, this equation leads to
%\small
%\begin{align}
%\mathrm d f(x_1^*,-\beta)/\mathrm d x +\mathrm d  f(x_{12}^*,\beta)/\mathrm d x+ \mathrm d f(x_{13}^*,2\beta)/\mathrm d x  =& 0 \nonumber\\
% \vdots \, & \nonumber\\
%\mathrm d f(x_{(n-1)(n-3)}^*,-2\beta)/\mathrm d x+\mathrm d f(x_{(n-1)(n-2)}^*,-\beta) /\mathrm d x &  \nonumber\\
%+\mathrm d f(x_{(n-1)n}^*,\beta)\mathrm d x = & 0 \label{eq:dfx_nn-1}  \\
%\mathrm d f(x_{ n(n-2)}^*,-2\beta)/\mathrm d x +\mathrm d f(x_{n(n-1)}^*,-\beta) /\mathrm d x = & 0 \label{eq:dfx_nn}  
%\end{align}
%\normalsize
% Since $x_{i(i-1)}^*\in [-\alpha_l,-\alpha_s]$, $x^*_{(n-1)(n-3)}\in [-2\alpha_l,-2\alpha_s]$. Then by \eqref{eq:eps1}, the first term of the left side of \eqref{eq:dfx_nn-1} should be within $[-\eps_{\mathcal I},\eps_{\mathcal I}]$. Moreover, by \eqref{eq:delta1}, \eqref{eq:delta2} and Assumption \ref{Asmp: f(p)diff2}, the second and third terms of the left side of equality \eqref{eq:dfx_nn-1} are less than $\delta_1$ and $\delta_2$, respectively. However, by the condition $\delta_2<-\delta_1-\eps_{\mathcal I}$, we have $\eps_{\mathcal I}+\delta_1+\delta_2<0$, namely the left side of equality \eqref{eq:dfx_nn-1} cannot equal to zero. This contradicts the existence of $x^*$. \qedp

\begin{proposition}\label{prop:gcee2}
	For $n\ge 3$, $\Pcal=[-\alpha_l,-\alpha_s]$ with $0<\alpha_s\le \alpha_l$ and $\beta>0$, assume that $f(\cdot)$ satisfies Assumption 1 and \ref{asmp:monotonicitydfpbeta}. If $\max_{x\in -\Qcal (2\Pcal)}f_x(x,-\beta)<-\max_{x\in \Pcal }f_x(x,-\beta)-\eps_{2\Pcal}$, then there exists no NE $X^*\in \R^n$ with $x_{i(i-1)}^*\in \Pcal$ for each $i\in V$.
\end{proposition} 

%\emph{Proof.} Suppose a RNE $x^*$ exists,  then it should satisfy equation \eqref{eq:dfx_nn-1} and \eqref{eq:dfx_nn}. Similar to the proof of Theorem \ref{thm:noeo2_3agents} and Proposition \ref{prop:gcee1}, by \eqref{eq:dfx_nn} and the assumption that $[-\alpha'_l,-\alpha'_s]$ is the unique interval within $ [-\alpha_l,\alpha_s]$ such that \eqref{eq:<=eps} is satisfied, we have $x_{(n-1)n}^*\in [-\alpha'_l, -\alpha'_s]$. Then by \eqref{eq:delta3} and the same argument as in the proof of Proposition \ref{prop:gcee1}, we have that equation \eqref{eq:dfx_nn-1} cannot hold. \qedp

\begin{proposition}\label{prop:gcee3}
For $n\ge 3$, $\Pcal=[-\alpha_l,-\alpha_s]$ with $0<\alpha_s\le \alpha_l$ and $\beta>0$, assume that $f(\cdot)$ satisfies Assumption 1, \ref{asmp:monotonicitydfpbeta} and \ref{asm:dfp2beta}. If $\max_{x\in -\Qcal (\mathcal I)}f_x(x,-\beta)<-\max_{x\in \Pcal }f_x(x,-\beta)$ with $\mathcal I =[-2\alpha_l,-2\alpha]$, then there exists no NE $X^*\in \R^n$ with $x_{i(i-1)}^*\in \Pcal$ for each $i\in V$.
\end{proposition}

\section{Nonexistence of the CE of interest}\label{Sec:CE}
This section shows a simple condition on the inter-agent benefit function $f$ under which the CE of interest cannot exist. It is based on the intuition that if the sum of the benefit of any two agents $f(p_{ij})+f(p_{ji})$ that are from each other, decreases as the longitudinal distance between them increases, then agents being cohesive longitudinally will increase the total benefit $J(p)$. In particular, if $x_0=\cdots = x_n$, then the total benefit $J$ attains the maximum.
% Since the leader $0$ is fixed at the origin and the network of agents is connected, the $x$ coordinate of the follower agents should also be zero to maximize the total benefit.
However, in this situation $x_{i(i-1)}\notin \Pcal$ for any negative interval $\Pcal$. Hence, even all agents stop with the same $x_i$, it is not a CE of interest. By an analogous reasoning, if $f(p_{ij})+f(p_{ji})$ always increases as the longitudinal distance between two agents increases, then there would not exist the CE of interest too. Based on these intuitions, the following result can be obtained.
\begin{theorem}\label{thm:nonce}
	For $n\ge 2$, $\Pcal=[-\alpha_l,-\alpha_s]$ with $0<\alpha_s\le \alpha_l$ and $\beta>0$, assume $f(\cdot)$ satisfies Assumption 1(a). If there exists a positive $\underline \beta\le \beta$ such that
	\begin{align}\label{eq:nonexofCE}
	& |f_x(x,y)+f_x(-x,-y))|>0\\
	& \forall x\in (0, \alpha_l], |y|\in [\underline \beta, +\infty) \nonumber
	\end{align}
then there exists no CE $\bar X^*\in \R^n$ with $\bar x^*_{i(i-1)}\in \Pcal$ for each $i\in \V$.
\end{theorem} 

\emph{Proof.} Assume that there exists the positive $\underline \beta\le \beta$ such that inequality \eqref{eq:nonexofCE} holds for $f(\cdot)$. Then, suppose there exists a CE $\bar X^*\in \R^{n}$ satisfying \eqref{eq:cooperequi} and $\bar x_{i(i-1)}^*\in \Pcal=[-\alpha_l,-\alpha_s]$ for each $i\in V$. Recall that $\bar X^*$ should satisfy \eqref{eq:group_opt}.
Consider this equality for $i=n$ and notice $\mathcal N_{n}=\{n-1,n-2\}$ and $\bar p^*_i=[\bar x_i^*,-i\beta]$, we have
\begin{align}\label{eq:CEanalysis}
&f_x(\bar x_{n(n-2)}^*,-2\beta) -f_x(\bar x_{(n-2)n}^*,2\beta)\nonumber\\
+& f_x(\bar x_{n(n-1)}^*, -\beta)
-f_x(\bar x_{(n-1)n}^*, \beta)  =0 
\end{align}

Since $\bar x^*_{(n-2)n}=-\bar x^*_{n(n-2)}$ and $\bar x^*_{(n-1)n}=-\bar x^*_{n(n-1)}$, the left side of the equality of \eqref{eq:CEanalysis} can be written as,
\begin{align}
&-\left(f_x(-\bar x_{(n-2)n}^*,-2\beta) + f_x(\bar x_{(n-2)n}^*,2\beta)\right) \nonumber\\
&-\left(f_x(-\bar x_{(n-1)n}^*,-\beta) +f_x(\bar x_{(n-1)n}^*,\beta)\right) \nonumber
\end{align}

However, by $\bar x^*_{(n-1)(n-2)}, \bar x^*_{n(n-1)}\in\Pcal$ and condition \eqref{eq:nonexofCE}, both the parentheses above should be positive and negative simultaneously, hence the expression above cannot equal zero. This violates \eqref{eq:CEanalysis}. \qedp

%\begin{remark}
%The sign of the inequality in \eqref{eq:nonexofCE} is the same for all $x,y$ satisfying the condition there. The result is also true if condition \eqref{eq:nonexofCE} is replaced by $\frac{\partial }{\partial x}(f(x,y)+f(-x,-y))<(>)0$ for all $  x\in (0, \bar \alpha]$ and $y=-\beta,-2\beta$, where the "less than" inequality for $y=-\beta$ may be deducted from the condition $\delta_2<-\delta_1$ in Theorem \eqref{thm:noeo_3agents}.
%\end{remark}
	
\begin{remark}
  Since condition \eqref{eq:nonexofCE} is proposed for every $y$ with $|y|\in [\underline \beta,+\infty)$, it can also be used to show the non-existence of the cooperative equilibrium of interest for the situation where agents adjust relative position in both directions within proper intervals. 
  %For instance, if $\underline \beta<\beta$ and $\overline \beta >2\beta$ and agents are allowed to change positions in both directions, by modifing condition \eqref{eq:nonexofCE}, we may there does not exist the cooperative equilibrium, where the relative position of each two neighboring agents is within $[-\alpha_l,-\alpha_s]\times [-\beta-\beta_\Delta,-\beta+\beta_\Delta]$, with $\beta_\Delta=\min \{\beta-\underline \beta,\overline \beta-2\beta\}$.
\end{remark}

\begin{remark}
	A simple class of benefit functions that satisfy condition \eqref{eq:nonexofCE} is $f(p)=g(x)h(y)$, where $h(y)$ is positive and differentiable with continuous derivative for all $y\in \R$, and $g(x)$ with $x\in \R$ is differentiable with continuous derivative, symmetric about the origin and strictly increasing or decreasing as $|x|$ increases, e.g., $|x|$, $x^2$, $\frac{1}{x^2}$ and the standard Gaussian function.
\end{remark}

\section{An application to line migratory formation}\label{Sec:Application}

\begin{figure}[!t]
	\centering
	\includegraphics[scale = 0.15]{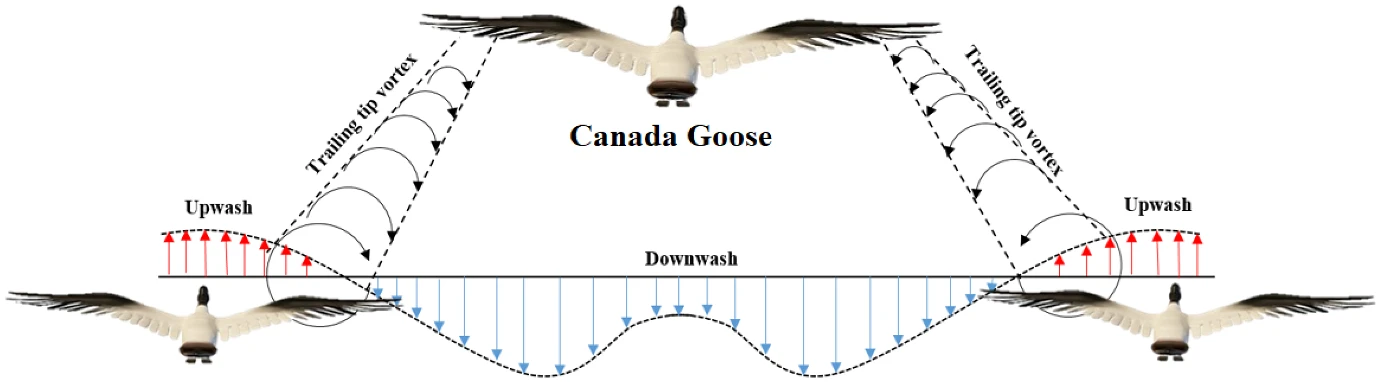}
	\caption{The movement of airflow around a bird.}
	\label{fig:horseshoe}
\end{figure}

In this section, we apply the theoretical results above to analyzing the emergence of the line formation of migrating birds. In most of existing researches on this topic, each bird is approximated by a fixed wing, whose forward motion stirs the air around upward and downward. If a bird positions properly relative to another bird, it can get extra lift from the upward airflow generated by that bird and reduce the energy used to counter the gravity, see Fig. \ref{fig:horseshoe}. This can be regarded as the wake benefit from one bird to another. 
The movement of the stirred air is usually depicted by a horseshoe vortex model \cite{hummel1983aerodynamic}, 
 Readers can refer to \cite{hummel1983aerodynamic} for more details on the model. We only introduce how to get the wake benefit here.

Assume that two birds $i=0,1$, with the same weight $W$ and wingspan $2b$ (the length of the wing), fly together along the $x$ direction with constant speed $U$, in the plane.
%The wake consists of a finite bound vortex with  circulation $\Gamma=\frac{W}{2\rho a U }$ and length $2a=\frac{\pi}{2}b$, where $\rho$ is the air density, and two semi-infinite trailing vortices  starting at the wingtips with vortex circulation $\Gamma$. 
If bird $0$ is at the origin $[0\ 0]^\top$, the upward airflow velocity $v(x,y)$ at $[x\ y]^\top\in \R^2$ generated by bird 0 can be given as

\small
\setlength{\abovedisplayskip}{1pt}
\begin{align}
&v(x,y)= v_b(x,y)+v_t(x,y) \label{eq:vp}\\
&v_b=\text{\footnotesize{$\frac{\Gamma}{4\pi} \frac{x}{x^2+r_0^2}\left[\frac{y+a}{\sqrt{(y+a)^2+x^2+r_0^2}}-\frac{y-a}{\sqrt{(y-a)^2+x^2+r_0^2}}\right]$}}\nonumber\\
&v_t= \text{\footnotesize{$\frac{\Gamma}{4\pi}\frac{y-a}{(y-a)^2+R(x)}\left[1-\frac{x}{\sqrt{(y-a)^2+x^2+R(x)}}\right]$}}\nonumber\\
&~~~~~~\text{\footnotesize{$-\frac{\Gamma}{4\pi}\frac{y+a}{(y+a)^2+R(x)}\left[1-\frac{x}{\sqrt{(y+a)^2+x^2+R(x)}}\right]$}} \nonumber
\end{align}
\normalsize
\noindent where $a=\frac{\pi}{4}b$, $\rho$ is the air density, $\Gamma=W/(2\rho a U)$, $R(x)=r_0^2+D_f|x|/U$ with $r_{0}=0.04b$ and $D_f$ is a diffusion term to model wake dissipation when $|x|\rightarrow \infty$ \cite{Greene-appmodel,Cattivelli-Model}. We select $D_f=1.05\times 10^{-4}Ub$ such that $\sqrt {R(x)}$ increases from $0.04b$ to $0.1b$ when $|x|$ grows from $0$ to $80b$, fairly realistic for aircraft wake \cite{de2013aircraft}. The model is valid for a sufficiently long longitudinal distance that covers the range of distances of neighboring birds in migratory formation. Beyond that distance, it is not accurate due to wake instability. 

\begin{figure}[t]
	\leftskip -0.4cm
	\rightskip -1 cm
	\subfigure[]{\includegraphics[scale=0.335]{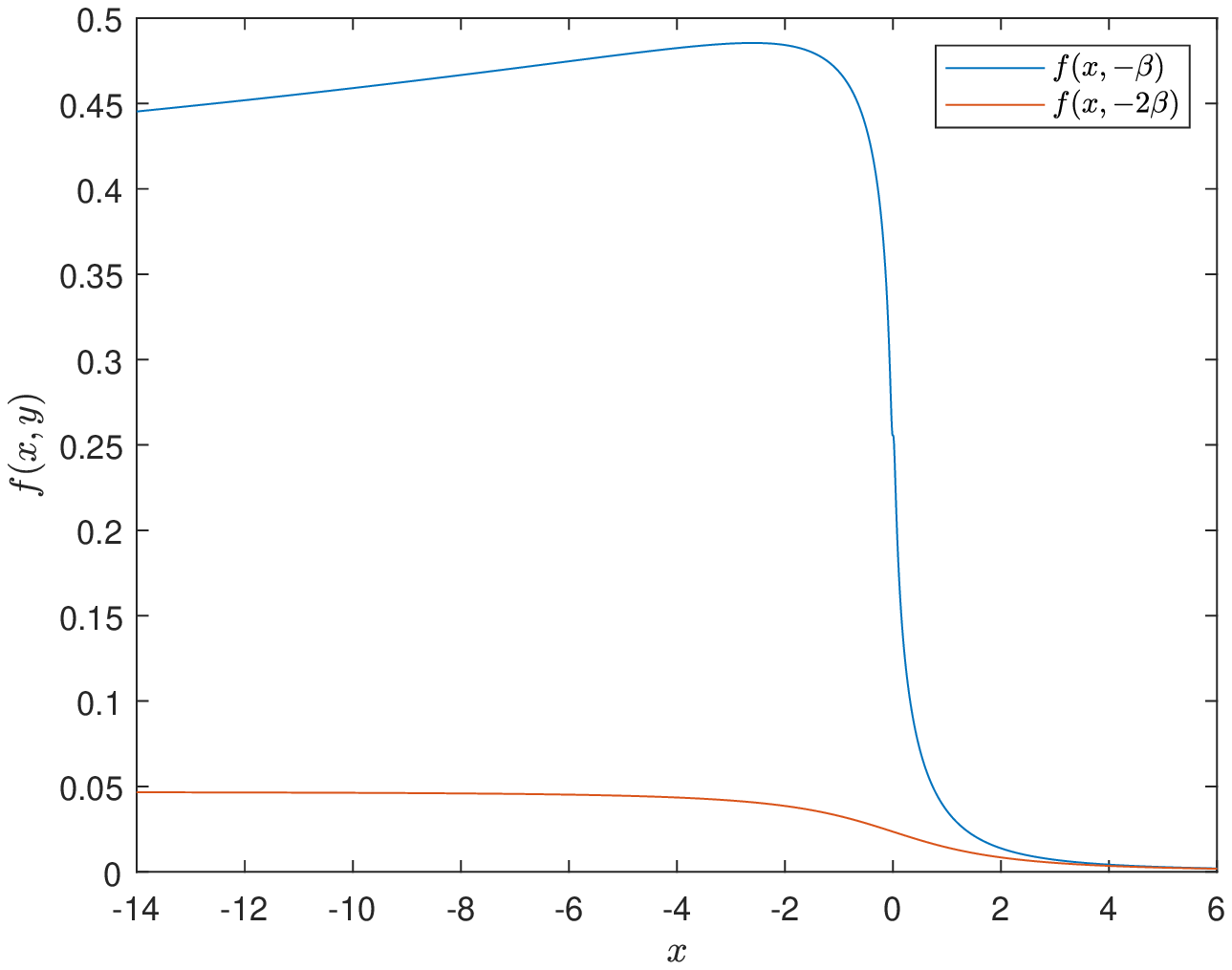}
		\label{fig:fp}	}
	\hspace{-0.825cm}
	\subfigure[]{\includegraphics[scale=0.335]{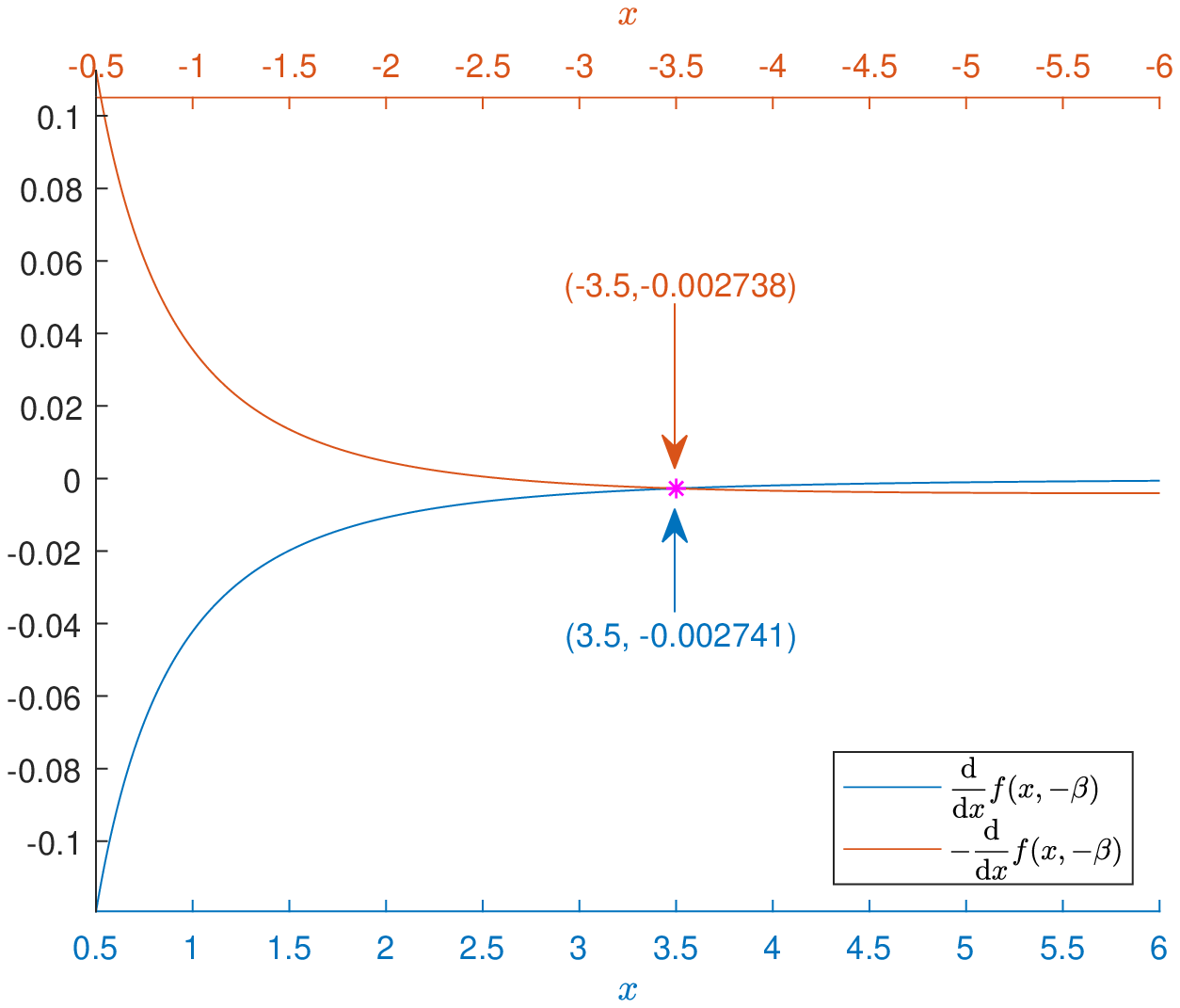}
		\label{fig:dfp_eofail_cond1}}
	\vspace*{-2mm}
	\caption{(a) The wake benefit $f(p)$ for $y=-\beta,-2\beta$ when $2b =1.5$ m. (b) The derivative of $f(x,-\beta)$. The point mark and $x$-axis with the same color corresponds to the derivative curve in that color. $\delta_2=-0.002741$ and $-\delta_1=-0.002738$.}
\end{figure}

Consider bird $1$ locating at $[x\ y]^\top$. After neglecting the momentum induced by the vertical airflow as in \cite{hummel1983aerodynamic}, the wake benefit of bird $1$ received from bird $0$ can be given as
\setlength{\abovedisplayskip}{5pt}
\begin{align}\label{eq:f_benefit}
f(x,y)= \frac{1}{2b}\int_{y-b}^{y+b} v(x,\eta) \mathrm{d} \eta 
\end{align}
This function satisfies Assumption 1 except for the points at the $y$ axis. Computation shows that $f(x,y)$ has a maximum $(-\alpha,-\beta)$ in the negative orthant, with $\alpha\approx 3.468 b$ and $\beta\approx (1+\frac{\pi}{4})b=a+b$, Moreover, it peaks around the line $y=-\beta$ in the negative orthant, which is argued \cite{Heppner_Orginizeflight} to be the best relative lateral position of a follower to its front neighbor. Hence, we fix $\beta$ in all the theorems previously as this value. Fig. \ref{fig:fp} shows $f(x,y)$ for $y=-\beta,-2\beta$ when $2b =1.5$. Normally, neighboring birds' longitudinal distance in migratory formation ranges from 0.5 wingspan ($1b$) to 4 wingspan ($8b$), implying $x_{i(i-1)}\in [-8b, -1b]$ in our setting.

As mentioned before, we have been working on reconstructing the line formation of migrating birds based on the assumption that birds behavior are purely guided by wake benefit maximization, taking into account birds attitudes. We have not numerically found the NE $X^*\in \R^n$ with $x^*_{i(i-1)}\in \Pcal$ and the CE $\bar X^*\in \R^n$ with $\bar x^*_{i(i-1)}\in \Pcal$ for a much wide interval $\Pcal$, e.g., $[-20b, -b]$ \cite{Mingming_unpublished}. In the following, we confirm this numerical result.

\begin{figure}[t]
	\leftskip -0.5cm
	\rightskip -1 cm
	\subfigure[$f_x(x,-\beta)$.]{
		\includegraphics[scale=0.33]{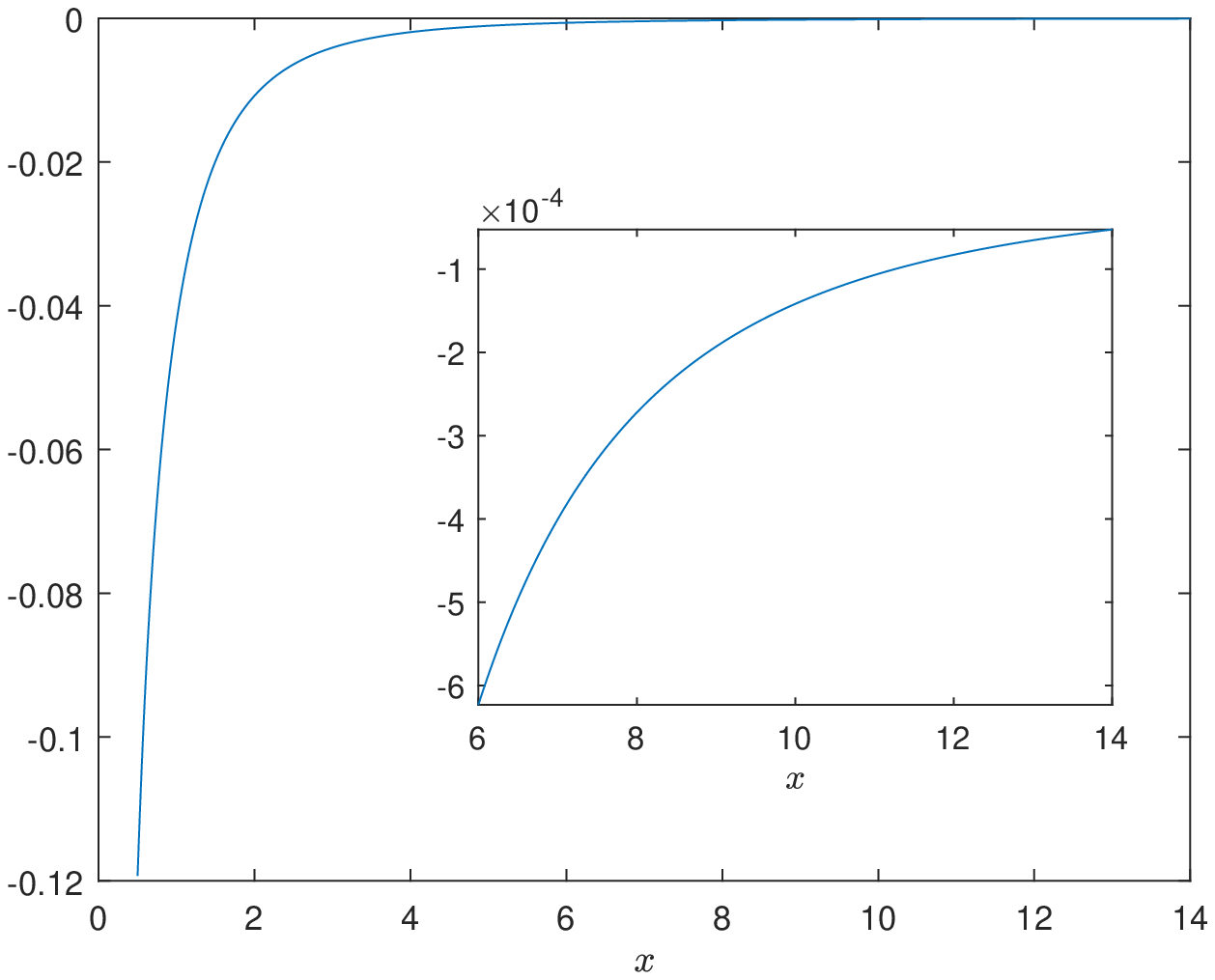}
		\label{fig:dfp_eofail_beta}	}
	\hspace{-0.9cm}
	\subfigure[$f_x(x,-2\beta)$.]{
		\includegraphics[scale=0.33]{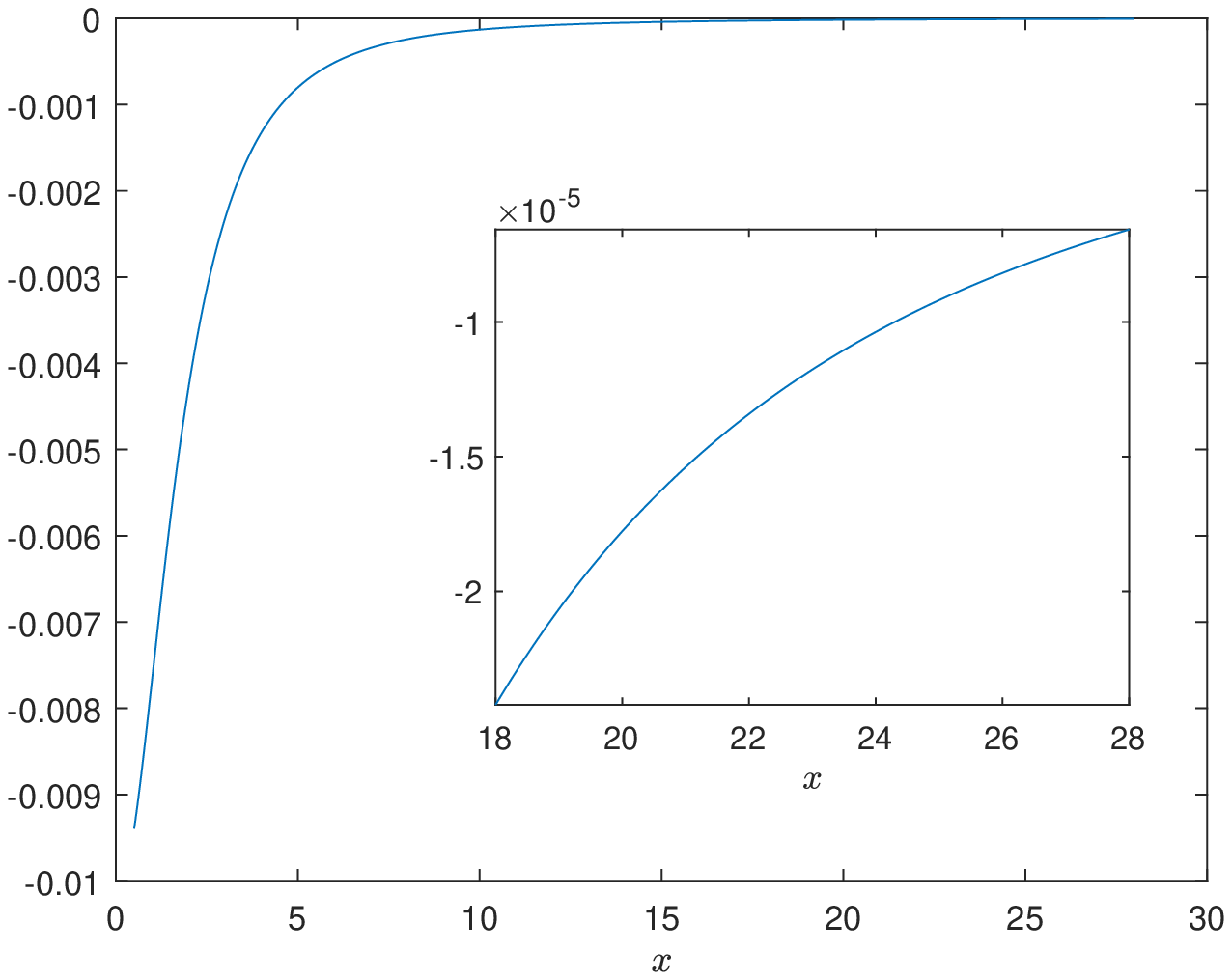}
		\label{fig:dfp_eofail_2beta}}
	\vspace*{-2mm}
	\caption{$f_x(x,y)$ for positive $x$.}
\end{figure}

\subsection{Absence of the NE of interest}

We first look at the selfish agents case. An illustrative example is presented for Canadian geese, which averagely have the weight $W=36.75$ N, wingspan $2b=1.5 $m, and fly with $U=18$ m/s at the height of $1$km from the ground during migration flight, where the air density $\rho\approx1.112\text{\ kg}/\text{m}^3$. Recall condition \eqref{eq:thm1} and \eqref{eq:thm2}, in the following, we always denote $\delta_1 =\max_{x\in \Pcal}f_x(x,-\beta)$, $\delta_2 =\max_{x\in -\Pcal }f_x(x,-\beta)$ and $\delta_3=\max_{x\in -\Qcal(\mathcal I)} f_x(x,-\beta)$ for the corresponding interval $\Pcal$ and $\Qcal(\mathcal I)$ accordingly.

By Fig. \ref{fig:dfp_eofail_cond1}, we can find that if we select $\Pcal= [-\alpha_l,-\alpha_s]=[-3.5,-0.5]$, $\delta_2\le -\delta_1$. Hence condition \eqref{eq:thm1} in Theorem \ref{thm:noeo_3agents} holds, and there should exist no NE $X^*$ with $x^*_{i(i-1)}\in \Pcal,i\in V$. Note that $\alpha_l$ cannot be increased more as the condition $\delta_2<-\delta_1$ would not be satisfied. 

\begin{figure}[t]
	\leftskip -0.5cm
	\rightskip -1 cm
	\subfigure[]{
		\includegraphics[scale=0.33]{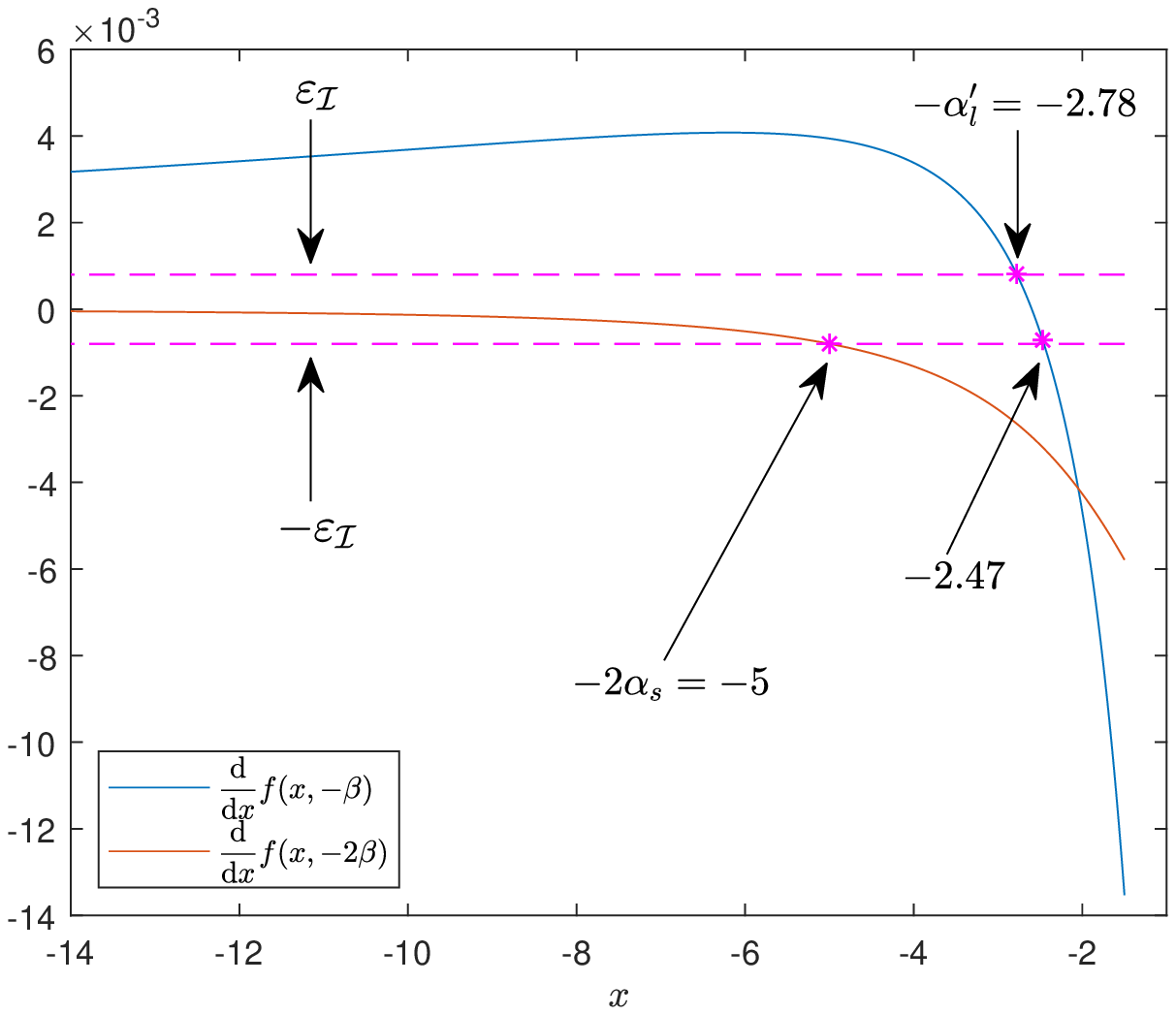}
		\label{fig:dfp_eofail_cond21}	}
	\hspace{-0.9cm}
	\subfigure[]{
		\includegraphics[scale=0.33]{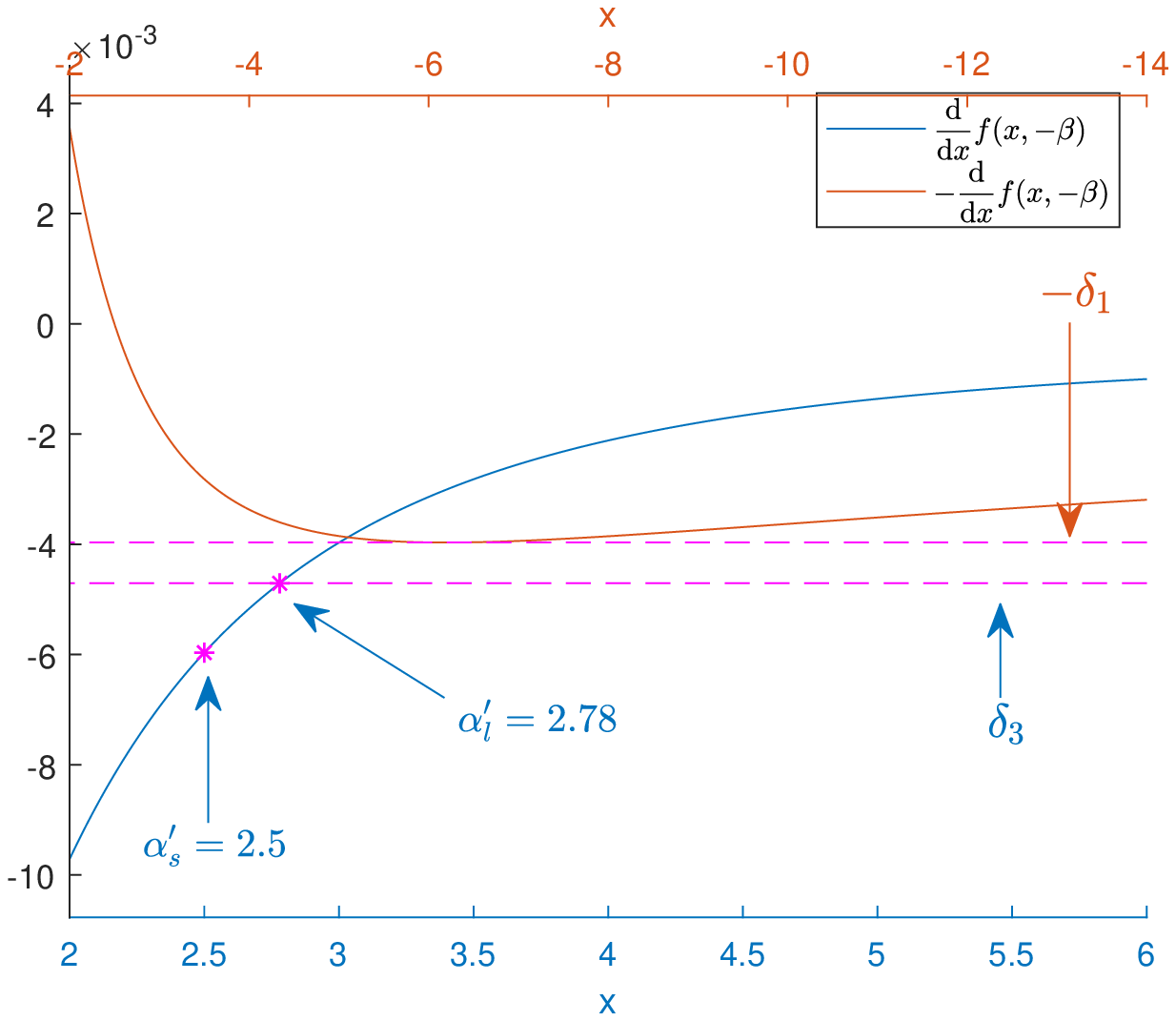}
		\label{fig:dfp_eofail_cond22}}
	\vspace*{-2mm}
	\caption{$f_x(x,y)$ for $y=-\beta$ and $-2\beta$. The value $\eps_{\mathcal I}=\max_{x\in \mathcal I}f_x(x,-2\beta)$ with ${\mathcal I}=2\Pcal=[-14,-5]$ is represented as the dashed magenta lines. The two ends of $\Qcal(\mathcal I)$ are $-\alpha'_l$ and $-\alpha'_s$.} 
\end{figure}

On the other hand, $f(x,-\beta)$ peaks at $-\alpha=-3.468b=-2.601$ and we can see from Fig. \ref{fig:dfp_eofail_beta} and \ref{fig:dfp_eofail_cond21} that Assumption \ref{asmp:monotonicitydfpbeta} is satisfied.  Let $\Pcal=[-\alpha_l,-\alpha_s]=[-7, -2.5]\ni -\alpha$, then ${\mathcal I}=2\Pcal=[-14,-5]$.  The value of $\eps_{\mathcal I}$ can be found in Fig. \ref{fig:dfp_eofail_cond21}. Moreover, the set $\Qcal(\mathcal I)=[-\alpha'_l,-\alpha'_s]$ satisfying \eqref{eq:<=eps} is a neighborhood of $-\alpha$ and given as $[-2.78,-2.5]$\footnote{Though the lower magenta line in Fig. \ref{fig:dfp_eofail_cond21} intersects $f_x(x,-\beta)$ at -2.47, $\Qcal=[-\alpha'_l,-\alpha'_s]$ should be the sub-set of $\Pcal$.}. Then by Fig. \ref{fig:dfp_eofail_cond22}, $\delta_3\le -\delta_1$ or condition \eqref{eq:thm2} holds. Hence, by Theorem \ref{thm:noeo2_3agents}, there exists no NE $X^*$ with $x^*_{i(i-1)}\in \Pcal,i\in V$ for $\Pcal=[-7, -2.5]$.
The value of $\alpha_s$ cannot be much smaller than 2.5, since from Fig. \ref{fig:dfp_eofail_cond21}, it would imply a wider $\mathcal I$, larger $\eps_{\mathcal I}$, wider interval $\Qcal(\mathcal I)$, and $\delta_3$ that would be larger than $-\delta_1$, making the condition in Theorem \ref{thm:noeo2_3agents} unsatisfied. 

\begin{figure}[!t]
	\leftskip -0.55cm
	\rightskip -1cm
	\subfigure[]{
		\includegraphics[scale=0.335]{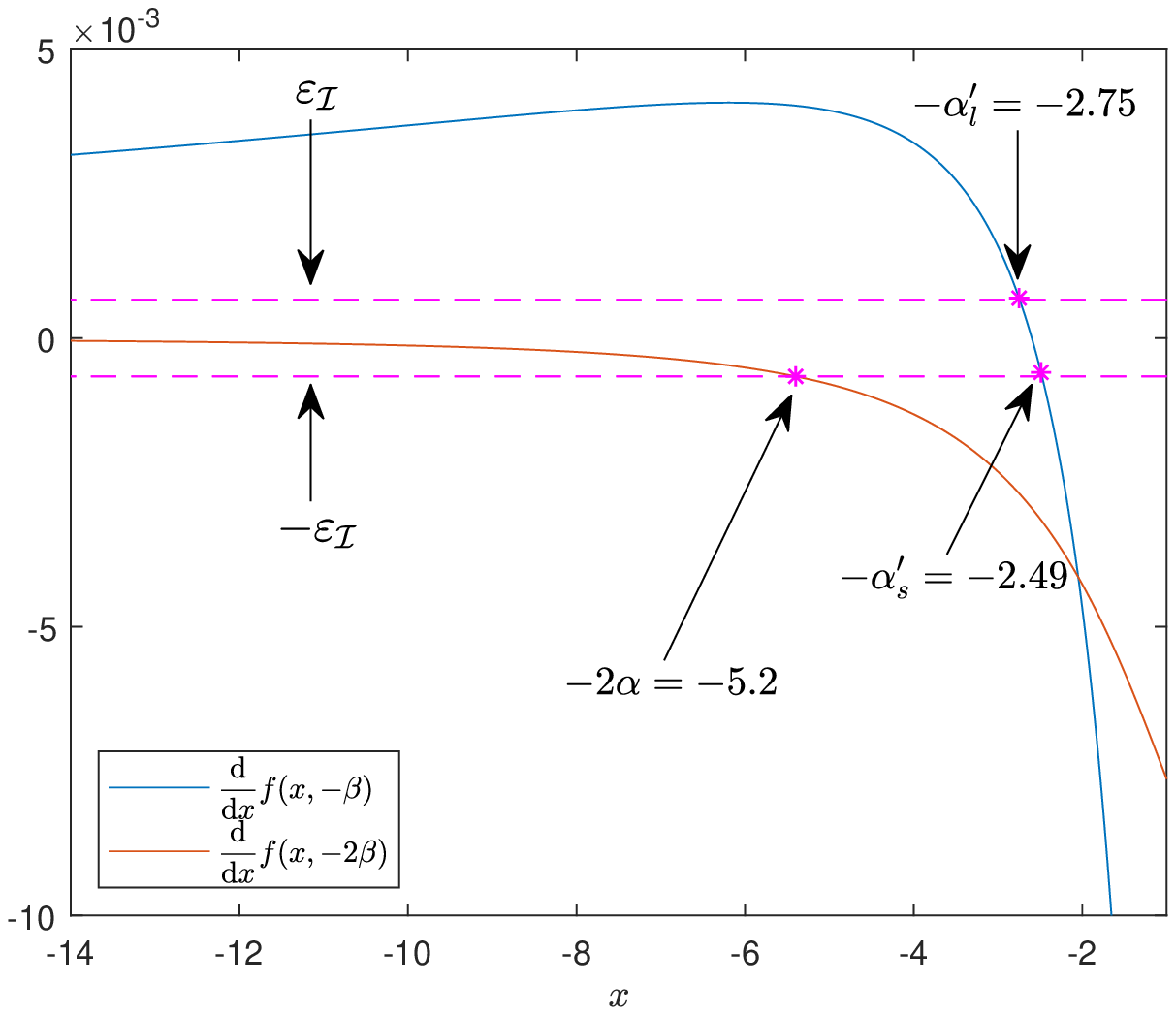}
		\label{fig:dfp_eofail_cond31}	}
	\hspace{-0.9cm}
	\subfigure[]{
		\includegraphics[scale=0.335]{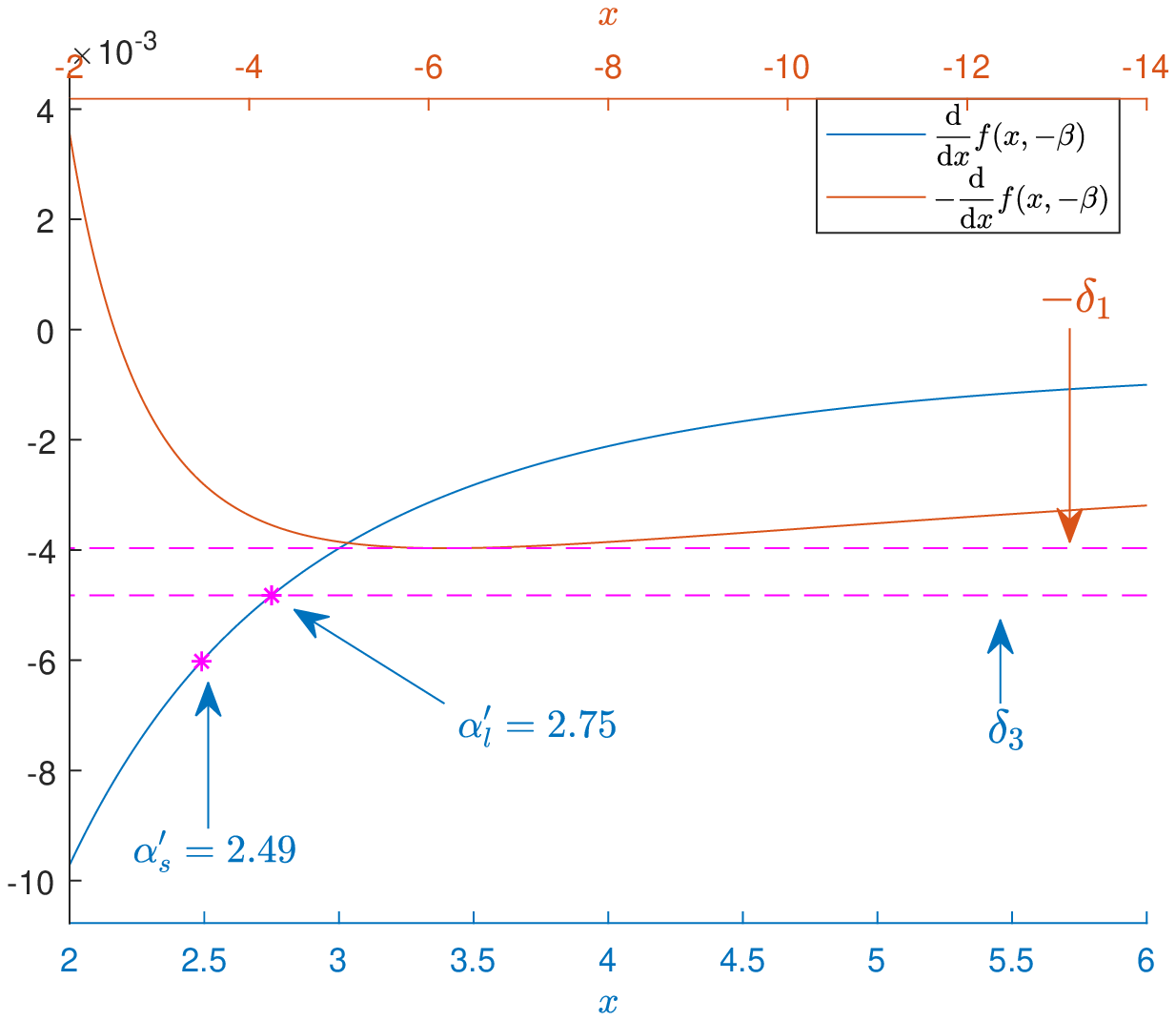}
		\label{fig:dfp_eofail_cond32}}
	\vspace*{-3mm}
	\caption{$f_x(x,y)$ for $y=-\beta$ and $-2\beta$. The value $\eps_{\mathcal I}$ with $\mathcal I=[-2\alpha_l,-2\alpha]=[-28,-5.2]$ is represented as the dashed magenta lines. The two ends of $\Qcal(\mathcal I)$ are $-\alpha'_l$ and $-\alpha'_s$.}
	\label{fig:dfp_eofail_cond3}
\end{figure}

Note as mentioned before this subsection, no NE $X^*$ of interest has been found for a wide $\Pcal$, e.g., $[-20b,-b]=[-15, -0.75]$. Hence, the two previous paragraphs are not able to explain this. We then turn to Theorem \ref{thm:noeo3_3agents}. By Fig. \ref{fig:dfp_eofail_2beta} and \ref{fig:dfp_eofail_cond31}, Assumption \ref{asm:dfp2beta} is indeed satisfied for $x\ge -2\alpha$ (The curve of $f_x(x,-2\beta)<0$ for $x\le -14$ is not shown for a clear vision of $\eps_{\mathcal I}$). Let $\Pcal=[-14,-0.5]\ni -\alpha$, then $\mathcal I = [-28,-1]$. In Fig. \ref{fig:dfp_eofail_cond31}, we can find $\Qcal(\mathcal I)=[-\alpha'_l,-\alpha'_s]$ that satisfies \eqref{eq:<=eps}. Then by Fig. \ref{fig:dfp_eofail_cond32}, $\delta_3<-\delta_1$. Hence conditions \eqref{eq:thm2} in Theorem \ref{thm:noeo3_3agents} holds for $\Pcal=[-14,-0.5]$, from which, we should have that there exists no NE $X^*$ of interest for this interval. This indeed explains our numerical search of the NE of interest.

 \subsection{Absence the CE of interest}
We then show that there exists no CE $\bar X^*$ of interest for a negative closed interval $\Pcal$ with the wake benefit function \eqref{eq:f_benefit}. For this function, we have following result.
\begin{lemma}\label{lem: wf_co}
	The benefit $f(x,y)$ in \eqref{eq:f_benefit} satisfies condition \eqref{eq:nonexofCE} with $\alpha_l\le\frac{U}{D_f}(2ab-r_0^2), \underline \beta\in (\sqrt{a^2+b^2},\beta)$.
\end{lemma}

The proof of this result is given in Appendix. Accordingly, we have the following result.
\begin{proposition}
	Consider the function $f(x,y)$ in \eqref{eq:f_benefit} and $n\ge 2$, then there exists no CE $\bar X^*\in\R^n$ such that $\bar x_{i(i-1)}^*\in \Pcal$ for any $\Pcal =[-\alpha_l,-\alpha_s]$ with $0<\alpha_s\le \alpha_l\le \frac{U}{D_f}(2ab-r_0^2)$.
\end{proposition}

\emph{Proof.} By Lemma \ref{lem: wf_co}, we have that condition \eqref{eq:nonexofCE} is satisfied with $\underline \beta\in (\sqrt{a^2+b^2}, \beta)$ and $\alpha_l\le\frac{U}{D_f}(2ab-r_0^2)$. This combining with Theorem \ref{thm:nonce} shows that for any $\Pcal=[-\alpha_l,-\alpha_s]$ with $0<\alpha_s\le  \alpha_l\le  \frac{U}{D_f}(2ab-r_0^2)$, there exists no CE of interest. \qedp

%The physical meaning of the proposition is that no echelon or one-side V formation, where every two neighboring birds have lateral distance of $\beta$ and longitudinal distance within $(0,\bar \alpha]$, can emerge when birds are cooperative to maximize the total wake benefit of their group. 

Putting $D_f=1.05\times 10^{-4} Ub$, $a=\frac{\pi}{4}b$ and $r_0=0.04b$ into $\frac{U}{D_f}(2ab-r_0^2)$ yields $ \alpha_l\le 14945 b$ or $7472$ wingspan. Based on the employed wake benefit model \eqref{eq:f_benefit}, the proposition predicts that no CE $\bar X^*$ of interest for an closed interval $\Pcal \subset [-14945b, 0)$ exists. Hence the echelon formation where each neighboring birds have the lateral distance of $(\frac{1}{2}+\frac{\pi}{8})$ wingspan and longitudinal distances less than $7472$ wingspans cannot emerge, when birds are cooperative to maximize the total wake benefit of the flock.
It should be noticed that echelon formation with the longitudinal distance of neighboring birds larger than $7472$ wingspans is not practical, as birds never fly so far from each other in formation flight. Furthermore, the fixed wing wake model may not be valid for such large longitudinal distance.

\section{Conclusion}\label{Sec:conclusion}
In this paper, we focus the two-dimensional echelon formation of multi-agents that behave to maximize relative-position dependent benefits. All the agents can be either selfish to maximize its own benefit from others or cooperative to optimize the total benefit of the group. We discuss the conditions on the inter-agent benefit such that echelon formations cannot appear, no matter agents are selfish or cooperative. The theoretical conditions are employed to analyze the fixed-wing model that is usually used to study line formations of  migrating birds, and justify our failure in numerically reconstructing migratory formations. This shows that the emergence of this kind formation may not emerge if birds behavior in migration is purely guided by energy savings. 

Our results imply multiple possibilities for the emergence reason of the migratory formations. First, remember that we employ the fixed-wings to model birds and ignore the slow undulatory motion of birds wings, conventionally as in \cite{Heppner_Orginizeflight,weimerskirch2001energy}, a natural hypothesis is
that the wing-flapping of birds plays more important roles than expected. Nevertheless, fixed-wings are proper to represent the glide of birds in formation flight. Hence, a second hypothesis from our result is that non-aerodynamic factors, such as collision avoidance and vision enhancement \cite{Heppner_Orginizeflight} could also take parts in developing the migratory formation.  Finally, from the perspective of multi-agent control systems, more complex dynamics, the actual sensing and information processing ability of the bird, and the communication capacity among birds (for cooperative birds) may need to be considered to see if the current result would still hold.

\bibliographystyle{IEEEtran}
\bibliography{bibitemNECE}

\begin{thebibliography}{10}
\providecommand{\url}[1]{#1}
\csname url@samestyle\endcsname
\providecommand{\newblock}{\relax}
\providecommand{\bibinfo}[2]{#2}
\providecommand{\BIBentrySTDinterwordspacing}{\spaceskip=0pt\relax}
\providecommand{\BIBentryALTinterwordstretchfactor}{4}
\providecommand{\BIBentryALTinterwordspacing}{\spaceskip=\fontdimen2\font plus
\BIBentryALTinterwordstretchfactor\fontdimen3\font minus
  \fontdimen4\font\relax}
\providecommand{\BIBforeignlanguage}[2]{{%
\expandafter\ifx\csname l@#1\endcsname\relax
\typeout{** WARNING: IEEEtran.bst: No hyphenation pattern has been}%
\typeout{** loaded for the language `#1'. Using the pattern for}%
\typeout{** the default language instead.}%
\else
\language=\csname l@#1\endcsname
\fi
#2}}
\providecommand{\BIBdecl}{\relax}
\BIBdecl

\bibitem{oh2015survey}
K.-K. Oh, M.-C. Park, and H.-S. Ahn, ``A survey of multi-agent formation
  control,'' \emph{Automatica}, vol.~53, pp. 424--440, 2015.

\bibitem{dong2008cooperative}
W.~Dong and J.~A. Farrell, ``Cooperative control of multiple nonholonomic
  mobile agents,'' \emph{IEEE Transactions on Automatic Control}, vol.~53,
  no.~6, pp. 1434--1448, 2008.

\bibitem{ren2007distributed}
W.~Ren and E.~Atkins, ``Distributed multi-vehicle coordinated control via local
  information exchange,'' \emph{International Journal of Robust and Nonlinear
  Control: IFAC-Affiliated Journal}, vol.~17, no. 10-11, pp. 1002--1033, 2007.

\bibitem{anderson2008rigid}
B.~D. Anderson, C.~Yu, B.~Fidan, and J.~M. Hendrickx, ``Rigid graph control
  architectures for autonomous formations,'' \emph{IEEE Control Systems
  Magazine}, vol.~28, no.~6, pp. 48--63, 2008.

\bibitem{hendrickx2007directed}
J.~M. Hendrickx, B.~D. Anderson, J.-C. Delvenne, and V.~D. Blondel, ``Directed
  graphs for the analysis of rigidity and persistence in autonomous agent
  systems,'' \emph{International Journal of Robust and Nonlinear Control:
  IFAC-Affiliated Journal}, vol.~17, no. 10-11, pp. 960--981, 2007.

\bibitem{zhao2015bearing}
S.~Zhao and D.~Zelazo, ``Bearing rigidity and almost global bearing-only
  formation stabilization,'' \emph{IEEE Transactions on Automatic Control},
  vol.~61, no.~5, pp. 1255--1268, 2015.

\bibitem{chen2020angle}
L.~Chen, M.~Cao, and C.~Li, ``Angle rigidity and its usage to stabilize
  multi-agent formations in 2d,'' \emph{IEEE Transactions on Automatic
  Control}, 2020.

\bibitem{reynolds1987flocks}
C.~W. Reynolds, ``Flocks, herds and schools: A distributed behavioral model,''
  in \emph{Proceedings of the 14th annual conference on Computer graphics and
  interactive techniques}, 1987, pp. 25--34.

\bibitem{vicsek1995novel}
T.~Vicsek, A.~Czir{\'o}k, E.~Ben-Jacob, I.~Cohen, and O.~Shochet, ``Novel type
  of phase transition in a system of self-driven particles,'' \emph{Physical
  review letters}, vol.~75, no.~6, p. 1226, 1995.

\bibitem{olfati2006flocking}
R.~Olfati-Saber, ``Flocking for multi-agent dynamic systems: Algorithms and
  theory,'' \emph{IEEE Transactions on automatic control}, vol.~51, no.~3, pp.
  401--420, 2006.

\bibitem{tanner2007flocking}
H.~G. Tanner, A.~Jadbabaie, and G.~J. Pappas, ``Flocking in fixed and switching
  networks,'' \emph{IEEE Transactions on Automatic control}, vol.~52, no.~5,
  pp. 863--868, 2007.

\bibitem{trenchard2016energy}
H.~Trenchard and M.~Perc, ``Energy saving mechanisms, collective behavior and
  the variation range hypothesis in biological systems: a review,''
  \emph{BioSystems}, vol. 147, pp. 40--66, 2016.

\bibitem{Heppner_Orginizeflight}
I.~L. Bajec and F.~H. Heppner, ``Organized flight in birds,'' \emph{Animal
  Behaviour}, vol.~78, no.~4, pp. 777 -- 789, 2009.

\bibitem{weimerskirch2001energy}
H.~Weimerskirch, J.~Martin, Y.~Clerquin, P.~Alexandre, and S.~Jiraskova,
  ``Energy saving in flight formation,'' \emph{Nature}, vol. 413, no. 6857, pp.
  697--698, 2001.

\bibitem{hummel1983aerodynamic}
D.~Hummel, ``Aerodynamic aspects of formation flight in birds,'' \emph{Journal
  of theoretical biology}, vol. 104, no.~3, pp. 321--347, 1983.

\bibitem{badgerow1981energy}
J.~P. Badgerow and F.~R. Hainsworth, ``Energy savings through formation flight?
  a re-examination of the vee formation,'' \emph{Journal of Theoretical
  Biology}, vol.~93, no.~1, pp. 41--52, 1981.

\bibitem{hummel1995formation}
D.~Hummel, ``Formation flight as an energy-saving mechanism,'' \emph{Israel
  Journal of Ecology and Evolution}, vol.~41, no.~3, pp. 261--278, 1995.

\bibitem{cutts1994energy}
C.~Cutts and J.~Speakman, ``Energy savings in formation flight of pink-footed
  geese,'' \emph{Journal of experimental biology}, vol. 189, no.~1, pp.
  251--261, 1994.

\bibitem{Cattivelli-Model}
F.~S. {Cattivelli} and A.~H. {Sayed}, ``Modeling bird flight formations using
  diffusion adaptation,'' \emph{IEEE Transactions on Signal Processing},
  vol.~59, no.~5, pp. 2038--2051, 2011.

\bibitem{li2017v}
X.~Li, Y.~Tan, J.~Fu, and I.~Mareels, ``On {V}-shaped flight formation of bird
  flocks with visual communication constraints,'' in \emph{13th IEEE
  International Conference on Control \& Automation (ICCA)}.\hskip 1em plus
  0.5em minus 0.4em\relax IEEE, 2017, pp. 513--518.

\bibitem{Greene-appmodel}
G.~C. Greene, ``An approximate model of vortex decay in the atmosphere,''
  \emph{Journal of Aircraft}, vol.~23, no.~7, pp. 566--573, 1986.

\bibitem{Mingming_unpublished}
M.~Shi and J.~M. Hendrickx, ``Whose energy cost would birds like to save: a
  revisit of the migratory formation flight,'' \emph{In preparation}.

\bibitem{bacsar1998dynamic}
T.~Ba{\c{s}}ar and G.~J. Olsder, \emph{Dynamic noncooperative game
  theory}.\hskip 1em plus 0.5em minus 0.4em\relax SIAM, 1998.

\bibitem{rosen1965existence}
J.~B. Rosen, ``Existence and uniqueness of equilibrium points for concave
  n-person games,'' \emph{Econometrica: Journal of the Econometric Society},
  pp. 520--534, 1965.

\bibitem{de2013aircraft}
I.~De~Visscher, L.~Bricteux, and G.~Winckelmans, ``Aircraft vortices in stably
  stratified and weakly turbulent atmospheres: Simulation and modeling,''
  \emph{AIAA journal}, vol.~51, no.~3, pp. 551--566, 2013.

\end{thebibliography}

\section*{Appendix: Proof of Lemma \ref{lem: wf_co}}

By \eqref{eq:vp} and \eqref{eq:f_benefit}, we have
\small
\begin{align}
f(p)=\frac{1}{2b}\int_{y-b}^{y+b} v_b(x,\eta) \mathrm{d} \eta +\frac{1}{2b}\int_{y-b}^{y+b} v_t(x,\eta) \mathrm{d} \eta  \nonumber
\end{align}
\normalsize

For  the two integration in the equation above, we can obtain
\small
\begin{align}
 &\frac{1}{2b}\int_{y-b}^{y+b} v_b(x,\eta) \mathrm{d} \eta\nonumber\\
= &\text{\footnotesize{$\frac{\Gamma}{8\pi b} \frac{x}{x^2+r_0^2}\left.\left[\sqrt{(\eta+a)^2+x^2+r_0^2}-\sqrt{(\eta-a)^2+x^2+r_0^2}\right]\right\vert^{y+b}_{y-b}$}}\nonumber\\
=& \frac{\Gamma}{8\pi b} \frac{x}{x^2+r_0^2}\left[\sqrt{c_3+x^2+r_0^2}-\sqrt{(c_2+x^2+r_0^2} \right.\nonumber\\
& ~~~~~~~~~~~~~~~~\left.-\sqrt{c_1+x^2+r_0^2}+\sqrt{c_4+x^2+r_0^2}\right]\label{eq:intvb}\\
&\frac{1}{2b} \int_{y-b}^{y+b} v_t(x,\eta) \mathrm{d} \eta \nonumber\\
 =&  \text{\footnotesize{$\frac{\Gamma}{8\pi b}\left[\ln \sqrt{(\eta-a)^2+R}-\frac{1}{2}\ln\left(\frac{\sqrt{(\eta-a)^2+x^2+R}-x}{\sqrt{(\eta-a)^2+x^2+R}+x}\right)\right.$}}\nonumber\\
& \text{\footnotesize{$\left.\left.- \sqrt{(\eta+a)^2+R}-\frac{1}{2}\ln\left(\frac{\sqrt{(\eta+a)^2+x^2+R}-x}{\sqrt{(\eta+a)^2+x^2+R}+x}\right)\right]\right\vert_{y-b}^{y+b}$}}\nonumber\\
=&  \frac{\Gamma}{8\pi b}f_{t1}(x,y)+\frac{\Gamma}{16\pi b}f_{t2}(x,y) \label{eq:intvt}
\end{align}
\normalsize
where
\small
\begin{align}
c_1=& (y+b-a)^2, \quad c_2=(y-b+a)^2\nonumber\\ 
c_3=& (y+b+a)^2,\quad c_4=(y-b-a)^2\nonumber\\
f_{t_1}(x,y)=& \frac{1}{2} \ln \frac{(c_1+R)(c_2+R)}{(c_3+R)(c_4+R)}\nonumber\\
f_{t_2}(x,y) =& \text{\footnotesize{$\ln\left(\frac{\sqrt{c_3+x^2+R}-x}{\sqrt{c_3+x^2+R}+x}\right)+\ln\left(\frac{\sqrt{c_4+x^2+R}-x}{\sqrt{c_4+x^2+R}+x}\right)$}}\nonumber\\
&\text{\footnotesize{$-\ln\left(\frac{\sqrt{c_1+x^2+R}-x}{\sqrt{c_1+x^2+R}+x}\right)-\ln\left(\frac{\sqrt{c_2+x^2+R}-x}{\sqrt{c_2+x^2+R}+x}\right)$}}\nonumber
\end{align}
\normalsize
%\begin{align}
% f_{t_1}(x,y)= & 
% \ln \sqrt{(y+b-a)^2+r_c^2(x)}-\ln \sqrt{(y-b-a)^2+r_c^2(x)} -\ln\sqrt{(y+b+a)^2+r_c^2(x)}+\ln \sqrt{(y-b+a)^2+r_c^2(x)}\nonumber\\
%f_{t_2}(x,y) =& \frac{1}{2}\left[\ln\left(\frac{\sqrt{(y+b+a)^2+x^2+r_c^2(x)}-x}{\sqrt{(y+b+a)^2+x^2+r_c^2(x)}+x}\right)-\ln\left(\frac{\sqrt{(y-b+a)^2+x^2+r_c^2(x)}-x}{\sqrt{(y-b+a)^2+x^2+r_c^2(x)}+x}\right)\right.\nonumber\\
%&\qquad\ \left.-\ln\left(\frac{\sqrt{(y+b-a)^2+x^2+r_c^2(x)}-x}{\sqrt{(y+b-a)^2+x^2+r_c^2(x)}+x}\right)
%+\ln\left(\frac{\sqrt{(y-b-a)^2+x^2+r_c^2(x)}-x}{\sqrt{(y-b-a)^2+x^2+r_c^2(x)}+x}\right)\right]\nonumber
%\end{align}
It should be notified that $c_1,c_2,c_3$, and $c_4$ depend on $y$. By equation \eqref{eq:intvb} and \eqref{eq:intvt}, one can check that
\small
\begin{align}
&\frac{1}{2b}\int_{y-b}^{y+b} v_b(x,\eta) \mathrm{d} \eta +\frac{1}{2b}\int_{-y-b}^{-y+b} v_b(-x,\eta) \mathrm{d} \eta=0, \nonumber\\
&f_{t1}(x,y)+f_{t1}(-x,-y)=\ln \frac{(c_1+R)(c_2+R)}{(c_3+R)(c_4+R)} \label{eq:ft1+-xy},\\
& f_{t2}(x,y)+f(-x,-y)=0. \nonumber
\end{align}
\normalsize
Then 
\small
\begin{align}
&f(x,y)+f(-x,-y) \nonumber\\
= &\frac{1}{2b}\int_{y-b}^{y+b} v_b(x,\eta) \mathrm{d} \eta +\frac{1}{2b}\int_{y-b}^{y+b} v_t(x,\eta) \mathrm{d} \eta  \nonumber\\
&+\frac{1}{2b}\int_{-y-b}^{-y+b} v_b(-x,\eta) \mathrm{d} \eta +\frac{1}{2b}\int_{-y-b}^{-y+b} v_t(-x,\eta) \mathrm{d} \eta \nonumber\\
=&\frac{1}{2b}\int_{y-b}^{y+b} v_t(x,\eta) \mathrm{d} \eta+\frac{1}{2b}\int_{-y-b}^{-y+b} v_t(-x,\eta) \mathrm{d} \eta\nonumber\\
=& \frac{\Gamma}{8\pi b}(f_{t1}(x,y)+f_{t1}(-x,-y))\nonumber\\
&+\frac{\Gamma}{16\pi b}(f_{t2}(x,y)+f_{t2}(-x,-y))\nonumber\\
=& \frac{\Gamma}{8\pi b}(f_{t1}(x,y)+f_{t1}(-x,-y))\nonumber
\end{align}
\normalsize
which according to \eqref{eq:ft1+-xy} is a function of $R$. Recall that $R(x)=r_0^2+\frac{D_f}{U}|x|=r_0^2+\frac{D_f}{U}x$ when $x\ge 0$,  we have 
\small
\begin{align}\label{eq:dft1(+-xy)}
&\frac{\partial }{\partial x}(f(x,y)+f(-x,-y)) \nonumber\\
=&\frac{\Gamma}{8\pi b}\frac{\partial}{\partial R}\left(f_{t1}(x,y)+f_{t1}(-x,-y)\right)\frac{\rmd R}{\rmd x}\nonumber\\
=& \frac{\Gamma}{8\pi b}\frac{\rmd R}{\rmd x} \frac{\partial }{\partial R} \ln \frac{(c_1+R)(c_2+R)}{(c_3+R)(c_4+R)}\nonumber\\
=& \frac{\Gamma}{8\pi b}\frac{\rmd R}{\rmd x} \frac{g(R)}{(c_1+R)(c_2+R)(c_3+R)(c_4+R)},
\end{align} 
\normalsize
where
\small
\begin{align}
g(R)=&(2R+c_1+c_2)(c_3+R)(c_4+R)\nonumber\\
&-(2R+c_3+c_4)(c_1+R)(c_2+R)\nonumber\\
%=& ((c_3+c_4)-(c_1+c_2))R^2+2(c_3c_4-c_1c_2)R\nonumber\\
%&+(c_1+c_2)c_3c_4-(c_3+c_4)c_1c_2\nonumber\\
=& 8ab(R^2+c_6R+c_7) \nonumber
\end{align}
\normalsize
with $c_6 = 2(a^2+b^2-y^2)$ and $c_7  = -3y^4+2(a^2+b^2)y^2+(a^2-b^2)^2$.

Note that when $x> 0$, $\frac{dR(x)}{dx}>0$, $R(x)>0$, and the denominator under $g(R)$ in \eqref{eq:dft1(+-xy)} is always positive, hence the sign of \eqref{eq:dft1(+-xy)} is the same as that of $g(R)$.
In order to check if there exists some positive $\underline \beta\le \beta$ such that condition \eqref{eq:nonexofCE} holds for the $\alpha_l$ satisfying the condition in the lemma, we test that if there exists this $\underline \beta$ such that for any $y$, with $|y|\in[\underline \beta,+\infty)$, the quadratic inequality $g(R)< 0$ holds when $R\in (r_0^2,r_0^2+\frac{D_f\alpha_l}{U}]$. It is easy to know that if $|y|\ge \sqrt{a^2+b^2}$, $c_6, c_7<0$
and $g(R)<0$ hold when
\small
\begin{align}
R\in& (0, \frac{-c_6+\sqrt{c_6^2-4c_7}}{2}) \nonumber\\
&= (0,y^2-(a^2+b^2)+2\sqrt{(y^2-a^2)(y^2-b^2)})\label{eq:interval}
\end{align} 
\normalsize
The right open end of this interval is an increasing function of $y^2$ when $|y|\ge \sqrt{a^2+b^2}$ and equals to $2ab$ when $|y|=\sqrt{a^2+b^2}$. Note that $\beta=a+b>\sqrt{a^2+b^2}$, if we take $\underline \beta\in (\sqrt{a^2+b^2},\beta)$, the intersection of the interval \eqref{eq:interval}  for all $|y|\in [ \underline \beta, +\infty)\subset [ \sqrt{a^2+b^2}, +\infty)$ is $(0,2ab+o)$ for some $o>0$. Hence, $g(R)<0$ holds for any $R\in (0,2ab]$ when $|y|\in [\underline \beta,+\infty)$. 

Recall $r_0^2=(0.02b)^2$ and $a=\frac{\pi}{4}b$, we have $2ab>r_0^2$ and for any positive $\alpha_l\le \frac{U}{D_f}(2ab-r_0^2)$, $(r_0^2,r_0^2+\frac{D_f}{U}\alpha_l]\subset (0,2ab]$. This shows that when $|y|\in  [\underline \beta,+\infty)$, for any $R\in (r_0^2,r_0^2+\frac{D_f}{U}\alpha_l]$ with $0<\alpha_l \le \frac{U}{D_f}(2ab-r_0^2)$, $g(R)<0$ holds. In other words, condition \eqref{eq:nonexofCE} holds for $ \alpha_l \le \frac{U}{D_f}(2ab-r_0^2)$, $\underline \beta\in (\sqrt{a^2+b^2},\beta)$. \qedp

%%%%%%%%%%%%%%%%%%%%%%%%%%%%%%%%%%%
%%%%%%%%%%%%%%%%%%%%%%%%%%%%%%%%%%%%%%%
\end{document}